\documentclass[journal]{IEEEtran}
\usepackage{cite}
\usepackage{booktabs}
\usepackage{multirow}
\usepackage[utf8]{inputenc}
\usepackage[T1]{fontenc}
\usepackage{amsmath,amssymb,amsfonts}
\usepackage{algorithm}
\usepackage{graphicx}
\usepackage{textcomp}
\usepackage{xcolor}
\usepackage{makecell}
\usepackage{pifont}
\usepackage{url}
\usepackage{algpseudocode}
\usepackage{amsthm}
\usepackage[hidelinks]{hyperref} % keep hyperref last
\usepackage{orcidlink}

\newtheorem{theorem}{Theorem}

\DeclareRobustCommand{\myurl}[1]{\url{#1}}

\def\BibTeX{{\rm B\kern-.05em{\sc i\kern-.025em b}\kern-.08em T\kern-.1667em\lower.7ex\hbox{E}\kern-.125emX}}

\newcommand{\x}{\mathbf{x}}
\newcommand{\y}{\mathbf{y}}

\newcommand{\X}{\mathbf{X}}
\newcommand{\Y}{\mathbf{Y}}
\newcommand{\z}{\mathbf{z}}

\newcommand{\task}{\mathcal{T}}

\newcommand{\eye}[1]{\mathbf{I}_{#1}}

\def \SemL{{\mathcal{L}_\mathrm{S}}}

\def \Tside{{\mathrm{T}}}              % transmitter side label
\def \Rside{{\mathrm{R}}}              % receiver side label

\def \Tdim{{q_{\Tside}}}
\def \Tdimc{{\overline{q}_{\Tside}}}

\def \Tlatr{\mathbf{s}_{\Tside}}

\def \Rdim{{q_{\Rside}}}
\def \Rdimc{{\overline{q}_{\Rside}}}

\def \Rlatr{\mathbf{s}_{\Rside}}

\def \reg{{\gamma}}

\def \AlignMatL{{\mathbf{A}_\mathrm{L}}}   % supervised linear alignment operator
\def \AlignMatF{{\mathbf{A}_\mathrm{F}}}   % PPFE alignment operator

\def \Target{{\mathbf{E}}}                 % SIM training target matrix
\def \TargetT{{\mathbf{E}_{\Tside}}}              % TX-side target factor
\def \TargetR{{\mathbf{E}_{\Rside}}}              % RX-side target factor
\def \TargetS{{\mathbf{E}_s}}              % generic per-SIM target
\def \TargetTtil{{\widetilde{\mathbf{E}}_{\Tside}}}   % channel-corrected TX target (sym. eq.)
\def \TargetRtil{{\widetilde{\mathbf{E}}_{\Rside}}}   % channel-corrected RX target (sym. eq.)

\def \LmatS{{\mathbf{B}_s}}               % generic left loss matrix
\def \RmatS{{\mathbf{C}_s}}               % generic right loss matrix

\def \UH{{\mathbf{U}_H}}
\def \VH{{\mathbf{V}_H}}
\def \SigH{{\boldsymbol{\Sigma}_H}}

\def \UT{{\mathbf{U}_E}}
\def \VT{{\mathbf{V}_E}}
\def \SigT{{\boldsymbol{\Sigma}_E}}

\def \PseudoInv{{\mathbf{H}^\dagger}}      % channel pseudo-inverse (wave-domain target, disjoint asym eq.)
\def \ChannelEq{{\mathbf{Q}_H}}            % digital ZF/MMSE equalizer (baseband, mono eq.)

\def \wsSteps{{t_{\mathrm{ws}}}}           % warm-start Adam steps
\def \maxIter{{t_{\max}}}                  % max optimization iterations

\def \GT{{\mathbf{G}_{\Tside}}}        % TX-SIM transfer matrix
\def \GR{{\mathbf{G}_{\Rside}}}        % RX-SIM transfer matrix
\def \Gs{{\mathbf{G}_s}}                  % generic per-SIM matrix
\def \Zeff{{\mathbf{Z}}}                  % effective cascade: \GR H \GT

\def \wavelength{{\lambda}}

\def \H{\mathbf{H}}

\def\Nt{{{N}_{\Tside}}}
\def\Nr{{{N}_{\Rside}}}
\def\snr{{\text{SNR}}}

\def \nclusters{\kappa}
\def \nsamples{\varrho}

\def \phv{\boldsymbol{\xi}}              % phase-shift vector of a SIM, e.g. \phv_{\Tside}
\def \phs{\xi}                           % scalar phase shift
\def \phset{\Xi}                         % box of admissible phases
\def \nph{p}                             % number of phase elements, e.g. \nph_{\Tside}
\def \resp{\boldsymbol{\upsilon}}        % element-response vector of a layer
\def \txsig{\mathbf{c}}                  % signal at the TX antennas
\def \cmap{\psi}                         % complex-domain mapping (pre-whitening)
\def \setA{\mathcal{A}}                  % anchor set
\def \setS{\mathcal{S}}                  % semantic pilots
\def \setD{\mathcal{D}}                  % dataset
\def \anch{\alpha}                       % anchor sample index, e.g. \anch_1
\def \Fm{\mathbf{F}}                     % Parseval frame operators, e.g. \Fm_{\Tside}

\def \bsc{\beta}                         % optimal scaling factor
\def \lr{\eta}                           % learning rate of the (projected) gradient steps
\def \regeps{\varepsilon}                % regularization of the channel inverse
\def \sigm{\varsigma}                    % elementwise sigmoid

\def \SigHsq{{\overline{\boldsymbol{\Sigma}}_H^{1/2}}}
\def \SigHisq{{\overline{\boldsymbol{\Sigma}}_H^{-1/2}}}

\def \LossJ{\mathcal{L}_{\mathrm{J}}}
\def \LossTrain{\mathcal{L}_{\mathrm{train}}}
\def \LossSup{\mathcal{L}_{\mathrm{sup}}}
\def \nmse{\mathrm{NMSE}}

\def \lip{\Lambda}                       % Lipschitz constants, e.g. \lip_{\Tside}
\def \logit{\boldsymbol{\vartheta}}      % unconstrained phase logits
\def \innov{\mathbf{g}}                  % normalized descent direction (negative NMSE gradient)
\def \rmsnorm{\Omega}                    % RMS normalization operator
\def \stepw{\mathbf{w}}                  % learnable step sizes
\def \mom{\boldsymbol{\mu}}              % learnable momentum coefficients
\def \scal{\boldsymbol{\tau}}            % learnable per-coordinate scaling
\def \mombuf{\mathbf{m}}                 % momentum (EMA) buffers
\def \params{\Theta}                     % set of learnable parameters
\def \dsw{\chi}                          % deep-supervision weights
\def \dsratio{\varkappa}                    % first-to-last weight ratio of the deep supervision

\def \chpert{\boldsymbol{\Gamma}}        % channel perturbation of the RPM model
\newcommand{\W}{\mathbf{W}}
\newcommand{\Ups}{\mathbf{\Upsilon}}
\def \amplconst{{\phi}}

\begin{document}

\title{Over-the-Air Semantic Alignment\\ via Dual Stacked Intelligent Metasurfaces \vspace{.2cm} }

% Replace the placeholder ORCID iDs (0000-0000-0000-0000) with the real ones.
% Membership grades are all set to "Member": change each one to
% Graduate Student Member / Senior Member / Fellow as appropriate.
\author{
Mario Edoardo Pandolfo\,\orcidlink{0009-0006-6509-4425},
\IEEEmembership{Student Member, IEEE},
Kyriakos Stylianopoulos\,\orcidlink{0000-0002-6842-2002},
\IEEEmembership{Student Member, IEEE},\\
George C. Alexandropoulos\,\orcidlink{0000-0002-6587-1371},
\IEEEmembership{Senior Member, IEEE},
and Paolo Di Lorenzo\,\orcidlink{0000-0002-4130-3177},
\IEEEmembership{Senior Member, IEEE}

\thanks{
A preliminary version of this work was presented at the 2026 European Signal Processing Conference (EUSIPCO)~\cite{pandolfo2026over}.
This work has been supported by the Smart Networks and Services
Joint Undertaking projects 6G-DISAC and 6G-GOALS under
the European Union's Horizon Europe research and innovation programme
under Grant Agreement numbers 101139130 and 101139232, respectively.}%
\thanks{M. E. Pandolfo is with the DIAG and DIET Departments, Sapienza
University of Rome, Italy, and also with the National
Inter-University Consortium for Telecommunications (CNIT), Parma, Italy
(e-mail: marioedoardo.pandolfo@uniroma1.it).}%
\thanks{P. Di Lorenzo is with the DIET Department, Sapienza University of
Rome, 00184 Rome, Italy, and also with CNIT, Parma, Italy
(e-mail: paolo.dilorenzo@uniroma1.it).}%
\thanks{K. Stylianopoulos and G. C. Alexandropoulos are with the Department
of Informatics and Telecommunications, National and Kapodistrian University
of Athens, Greece (e-mails: \{kstylianop,alexandg\}@di.uoa.gr).} \vspace{-.6cm} }

\markboth{}{}

\maketitle

\begin{abstract}
Semantic communications enable task-oriented information exchange, but suffer from latent-space mismatch when transmitter and receiver employ heterogeneous machine learning models. Existing semantic alignment methods rely on digital processing, introducing computational and latency overhead. This paper proposes a dual stacked intelligent metasurfaces (dual-SIM) architecture that performs semantic alignment entirely in the wave domain. By jointly exploiting programmable metasurfaces at both ends of a link and the wireless channel, the proposed architecture realizes an end-to-end linear transformation between heterogeneous latent spaces. We show phase-only multilayer SIM can approximate both supervised linear aligners and zero-shot Parseval-frame equalizers. We investigate single-SIM, disjoint asymmetric and symmetric dual-SIM, and jointly optimized dual-SIM configurations, considering channel equalization and constructive channel exploitation. For joint design, we develop an alternating projected gradient algorithm with convergence guarantees to first-order stationary points. We further introduce a deep-unrolled network that unrolls gradient-based iterations to reduce computational cost of SIM reconfiguration under channel variations. Numerical results with heterogeneous vision transformer encoders show accurate semantic alignment from moderate to high signal-to-noise ratios. The joint and symmetric disjoint dual-SIM designs achieve the highest accuracy over compact channels, the joint design leads at shallow stack depths, and the advantage over single-SIM designs diminishes at equal depth.
\end{abstract}

\begin{IEEEkeywords}
Semantic communications, semantic equalization, stacked intelligent metasurfaces, latent space alignment, wave-domain processing. 
\end{IEEEkeywords}

\section{Introduction}

Conventional communication systems, rooted in Shannon's source--channel separation theorem, excel at reliable bit delivery but face increasingly stringent latency, bandwidth, and energy requirements from emerging applications, including autonomous systems, industrial automation, and the Internet of Things~\cite{alwis2021devices}. Semantic communications (SC) address these challenges by shifting the focus from exact bit recovery to the transmission of task-relevant information~\cite{strinati20216g, gunduz2022beyond, strinati2024goal, Yang2023SemComFutureInternetCST}. Deep neural networks (DNNs) typically extract compact, low-dimensional latent representations that replace conventional information streams, enabling communication at the level of meaning or task utility with reduced overhead and energy consumption.

Two representative SC paradigms are edge inference (EI) and deep joint source--channel coding (DJSCC). In EI, resource-constrained devices cooperate with edge servers to execute DNNs through distributed or split inference~\cite{deng2020edge}. In DJSCC, neural encoders and decoders are jointly trained to learn semantic compression and channel coding end to end~\cite{gunduz2024joint, bourtsoulatze2019deep, xu2023deep}. Despite their effectiveness, both EI and DJSCC typically assume that the transmitter (TX) and receiver (RX) share a common latent space. This assumption is often unrealistic in heterogeneous networks where devices may employ independently trained models, different architectures, or proprietary designs that cannot be exchanged due to privacy or intellectual-property constraints. The resulting latent-space mismatch introduces \textit{semantic noise}, potentially degrading task performance even over an ideal physical channel~\cite{luo2022semantic, sana2023semantic}. \emph{Semantic alignment}, also referred to as \emph{semantic channel equalization}, addresses this challenge by mapping latent representations across heterogeneous models, enabling interoperability without joint training or model exchange~\cite{sana2023semantic}.

\noindent\textbf{Related Works}. Existing approaches to cross-model alignment fall into two families. \textit{Supervised} methods learn linear or structured mappings between latent spaces~\cite{merullo2022linearly, moayeri2023text, maiorca2023latent, lahner2024direct}, using orthogonal Procrustes analysis~\cite{wang2008manifold}, communication-aware optimization~\cite{pandolfo2025latent,huttebraucker2025ris,di2025federated}, or DNN-based mappings for cross-model alignment in DJSCC systems~\cite{pannacci2025semantic,pandolfo2025latent}. These methods require exchanging \textit{semantic pilots} (SPs), i.e., paired latent representations or task-specific examples. In contrast, \textit{unsupervised zero-shot} methods avoid pilot exchange by exploiting \textit{anchors}, reference samples processed independently by each model. Anchor-based representations are invariant to rotations, reflections, and permutations~\cite{moschella2023relative}, enabling alignment when latent spaces approximately preserve their relative geometry. This principle has been applied to semantic channel equalization~\cite{huttebraucker2024relative} and extended to dynamic multi-agent settings~\cite{fiorellino2025frame}. However, architectural and training differences may introduce scaling and anisotropy beyond isometric transformations, motivating latent-space whitening~\cite{pandolfo2026semasia}. More recently, \cite{svendsen2026improving} incorporated whitening into an affine-invariant similarity measure, and introduced learnable anchors constructed from convex combinations of reference samples. Despite these advances, both families rely on digital pre- and/or post-alignment stages, increasing computational complexity and latency. This motivates shifting semantic alignment toward the radio hardware.

Stacked intelligent metasurfaces (SIM) have emerged as a platform for wave-domain computation~\cite{SIM_Survey}, leveraging cascaded programmable electromagnetic layers to realize high-dimensional linear transformations directly over the air (OTA), with greater spatial degrees of freedom than single-layer reconfigurable intelligent surfaces (RISs)~\cite{AlexandropoulosRIS} and reduced baseband processing. Early studies established their potential for holographic multiple-input multiple-output (MIMO), multiuser communications, and wave-domain signal processing~\cite{An2024SIMTransceiver_MWC,liu2025stacked,AXN23_SIM,An2025SIMMultiuserDownlinkTWC,10949617,11157883,SIM_UAVs}. In particular, the dual-SIM architecture, with jointly optimized metasurfaces at the transmitter (TX) and receiver (RX), was introduced for end-to-end channel shaping~\cite{AXN23_SIM} and subsequently extended to hybrid analog--digital designs~\cite{Perovic2024MISIMHMIMOCL}. Beyond channel optimization, SIMs have demonstrated the ability to implement structured operators, including the two-dimensional discrete Fourier transform (2D-DFT) for direction-of-arrival estimation under phase-only constraints~\cite{An2024SIM2DDFT_ICC,An2024SIM2DDFT_JSAC}. Other applications include near-field beamforming~\cite{Papazafeiropoulos2024NearFieldSIMWCL}, low Earth orbit satellite communications~\cite{Lin2024SIMLEOSatWCL}, secure transmission~\cite{Niu2025EfficientSIMSecureTIFS}, integrated communication and computation~\cite{Chen2026SIMICCIoTJ}, and interference-exploitation precoding~\cite{Wang2026SIMInterferenceFreePrecodingTWC}.

More recently, SIM research has expanded toward task-oriented and artificial intelligence (AI)-native communications. Early RIS-based diffractive neural networks implemented paired semantic encoders and decoders for data reconstruction~\cite{Chen2024RISOTASemComMWC}, while subsequent SIM-based designs optimized task performance~\cite{Huang2025SIMSemanticWCL}, supported multimodal semantic representations~\cite{Huang2025SIMMultiModalWCL}, and enabled distributed image-semantic encoding through SIM mounted on unmanned aerial vehicles~\cite{Li2026DistributedEMNNSemCom}. Hybrid architectures combining SIMs with electronic neural networks have also been investigated~\cite{Liu2025HOENNSemComICC}, alongside image-based evaluations of SIM-enabled semantic communications~\cite{An2026EmergingMetasurfacesTAP}. However, these approaches do not explicitly address latent-space mismatch between heterogeneous AI models. Our preliminary work~\cite{pandolfo2026over,stylianopoulos2026metasurfaces} tackled this problem using a single TX-side SIM followed by digital channel equalization at the RX. Related developments include metasurface-integrated neural networks (MINN)~\cite{Stylianopoulos2025OTAEI,stylianopoulos2026metasurfaces} and SIM-based extremely large MIMO systems~\cite{XL-MIMO-ELM} for wave-domain EI. %, as well as a hybrid analog--digital semantic equalization using a fabricated dynamic metasurface antenna~\cite{Sghaier2026WaveDomainSemanticEqDMA}. 
%Surveys of SIM-enabled systems are available in~\cite{liu2025stacked,Liu2026SIMReviewElectronics,An2026SIMWaveDomainSurveyCOMST,Shi2026SIM6GNetworksMWC}.

\noindent\textbf{Contributions}. This paper substantially extends our preliminary conference work~\cite{pandolfo2026over}, where OTA semantic alignment was first investigated using a single TX-side SIM followed by digital channel equalization at the RX. Here, we move beyond this architecture by introducing dual-SIM processing, where the two metasurfaces and the intervening wireless channel jointly realize the semantic alignment operator entirely in the wave domain. Unlike the subsequent independent work~\cite{Sghaier2026WaveDomainSemanticEqDMA}, which combines a fabricated dynamic metasurface antenna with digital pre- and post-equalizers, our framework targets the direct wave-domain realization of the complete alignment operator through cascaded multilayer SIMs at both ends of the link. The main contributions of this journal extension are as follows.

\begin{itemize}

\item \textit{Dual-SIM semantic alignment.}
We introduce a dual-SIM architecture for OTA semantic alignment between heterogeneous AI models, where multilayer metasurfaces at the TX and RX jointly realize an end-to-end alignment operator through the wireless channel. Extending our preliminary single-SIM design~\cite{pandolfo2026over}, the proposed architecture integrates semantic alignment and channel compensation into a single electromagnetic transformation, enabling fully OTA emulation of both supervised linear aligners and zero-shot Parseval-frame equalizers, without the need for dedicated digital equalization stages.

\item \textit{Unified SIM deployment and design framework.}
We investigate three SIM deployment strategies: single-SIM, disjoint dual-SIM with asymmetric or symmetric designs, and jointly optimized dual-SIM. For each applicable configuration, we consider channel equalization or constructive channel exploitation within the alignment process. We characterize their trade-offs in hardware complexity, channel knowledge, and TX--RX coordination. Interestingly, unlike single-SIM and asymmetric designs, symmetric and joint configurations also support transmission over fewer antennas than latent symbols, subject to the rank constraints of the effective channel.

\item \textit{Joint physical--semantic optimization.}
We formulate the joint dual-SIM design as a hardware-constrained inverse problem to approximate the target semantic operator through the cascaded electromagnetic transformation under phase-only multilayer metasurface constraints. We derive an alternating projected gradient algorithm with closed-form scalar calibration, establishing convergence of objective values and first-order stationarity of limit points under suitable assumptions. This extends the single-SIM gradient-based formulation of~\cite{pandolfo2026over} to joint optimization of the physical--semantic cascade.

\item \textit{Deep-unrolled SIM reconfiguration.}
Following the model-based deep learning paradigm~\cite{Shlezinger2023MBLearning,Balatsoukas2019DeepUnfolding}, we propose a deep-unrolled network that unrolls gradient-based iterations of the joint optimizer, with learnable per-coordinate step sizes and momentum in place of fixed steps. Trained offline on a channel distribution with a semantic warm start, it yields suitable configurations of both SIMs for unseen channels with few gradient evaluations, avoiding the cost of per-channel iterative optimization.

\end{itemize}

\begin{figure*}[t]
    \centering
    \includegraphics[width=\linewidth, trim=0bp 5bp 0bp 180bp, clip]{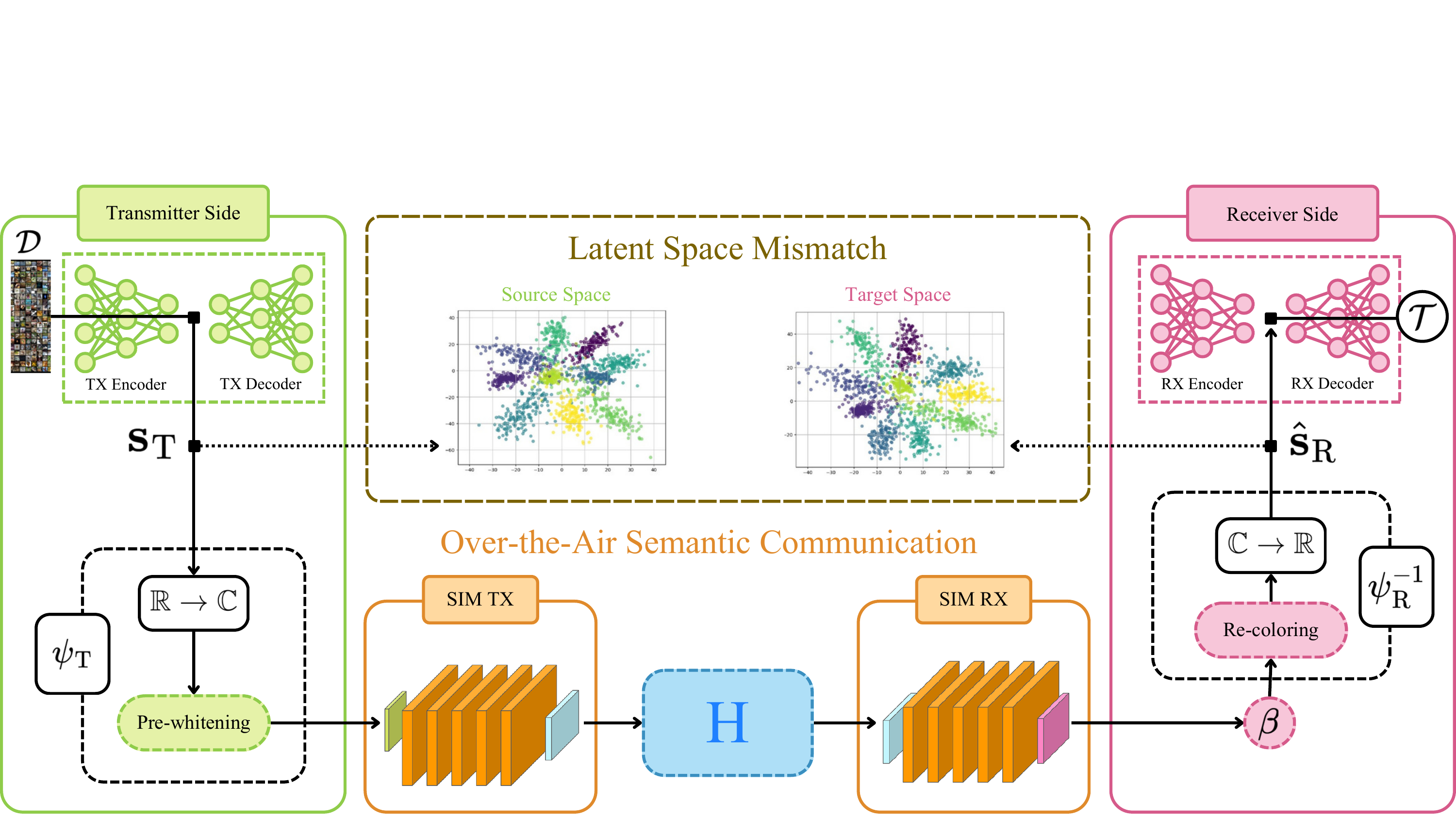}
    \caption{The considered dual-SIM semantic communications system incorporating stacked metasurfaces at both the TX and RX sides. The cascaded electromagnetic transformation $\GR \H \GT$ jointly performs physical-channel transformation and semantic alignment entirely OTA, i.e., in the wave domain.}
    \label{fig:system_model}
\end{figure*}

% \noindent\textbf{Outline.} Section~\ref{sec:system_model} presents the dual-SIM system model, while Section~\ref{sec:optimization} introduces supervised and zero-shot semantic alignment. Section~\ref{sec:sim_schemes} describes the SIM deployment schemes, and Section~\ref{sec:dual_optim} develops the joint optimization algorithm with convergence guarantees. Section~\ref{sec:deep_unrolling} introduces the deep-unrolled architecture. Section~\ref{sec:numerical} presents numerical results on system parameters, channel conditions, deep unrolling, and image reconstruction. Finally, Section~\ref{sec:conclusion} draws the conclusions of the paper.
\noindent\textbf{Outline.} Section~\ref{sec:system_model} presents our dual-SIM system model, while Section~\ref{sec:optimization} introduces supervised and zero-shot semantic alignment. Section~\ref{sec:sim_schemes} describes the SIM deployment schemes, and Section~\ref{sec:dual_optim} develops the joint optimization, with an alternating projected gradient algorithm with convergence guarantees in Sec.~\ref{sec:alt_pg} and the deep-unrolled architecture in Sec.~\ref{sec:deep_unrolling}. Section~\ref{sec:numerical} presents numerical results on system parameters, channel conditions, and deep unrolling. Finally, Section~\ref{sec:conclusion} draws the conclusions of the paper.

\noindent\textbf{Notation.}
Scalars, column vectors, and matrices are denoted by $a$, $\mathbf{a}$, and $\mathbf{A}$, respectively. Their entries are indexed as $[\mathbf{a}]_i$ and $[\mathbf{A}]_{i,j}$, while $\mathbf{I}$ is the identity matrix. $(\cdot)^H$ and $(\cdot)^\top$ denote Hermitian transpose and transpose, while $\mathrm{tr}(\cdot)$, $\|\cdot\|_F$, $\mathrm{vec}(\cdot)$, and $\mathrm{diag}(\cdot)$ denote trace, Frobenius norm, column-wise vectorization, and diagonal matrix of a vector, respectively. $\Re\{\cdot\}$ and $\Im\{\cdot\}$ denote real and imaginary parts, $\jmath\triangleq\sqrt{-1}$, and $\frac{\partial \mathbf{f}}{\partial \mathbf{x}^\top}$ is the Jacobian of $\mathbf{f}(\mathbf{x})$. Finally, $\Pi_{[0,2\pi)}(\cdot)$ is the elementwise projection onto $[0,2\pi)$, while $\odot$, $\oslash$, and $(\cdot)^{\circ (\cdot)}$ denote elementwise multiplication, division, and power.

\section{System Model}
\label{sec:system_model}

We consider a MIMO SC setting, as depicted in Fig.~\ref{fig:system_model}, in which a TX supports a downstream task $\task$ (e.g., classification) carried out at an RX. Each side runs its own \textit{pre-trained} neural network, and these two networks are \textit{heterogeneous}: they are developed independently and are neither trained jointly nor shared with the other party. For a data sample $\delta \in \setD$, the TX encoder outputs a semantic feature vector $\Tlatr \in \mathbb{R}^{\Tdim}$, and the collection of all such vectors constitutes the TX latent semantic space. At the RX, $\Rlatr \in \mathbb{R}^{\Rdim}$ denotes the representation of the same sample that the RX decoder expects in order to correctly perform $\task$, and the collection of these vectors constitutes the RX latent semantic space. Since the two networks have different architectures, their latent spaces generally differ in both structure and dimensionality. In fact, directly feeding TX features to the RX is possible only when their dimensions match, but even then, the representation mismatch introduces \emph{semantic noise} that degrades performance on $\task$. Semantic alignment is thus needed to bridge the two latent spaces.

To enable wave-domain processing through the SIM attached at the end of the link, the real-valued latent representations are first mapped into complex-valued baseband symbols. Assuming, without loss of generality, that $\Tdim$ and $\Rdim$ are even, each complex symbol is built from one entry of the first half of the feature vector and one entry of the second half. A pre-whitening step then removes the individual scale and anisotropy of each latent space~\cite{pandolfo2026semasia}, and makes the symbols better suited to wave-domain processing. We denote the resulting transformation by $\cmap_{\Tside}:\mathbb{R}^{\Tdim}\to\mathbb{C}^{\Tdimc}$ at the TX and $\cmap_{\Rside}:\mathbb{R}^{\Rdim}\to\mathbb{C}^{\Rdimc}$ at the RX, where $\Tdimc \triangleq \Tdim/2$ and $\Rdimc \triangleq \Rdim/2$, since the two architectures differ. The complex pre-whitened vectors are represented as follows:
\begin{equation}
    \x \triangleq \cmap_{\Tside}(\Tlatr) \in \mathbb{C}^{\Tdimc},
\qquad
\y \triangleq \cmap_{\Rside}(\Rlatr) \in \mathbb{C}^{\Rdimc}.
\end{equation}
The TX-side SIM implements the parameterized electromagnetic transformation $\GT(\phv_{\Tside}) \in \mathbb{C}^{\Nt \times \Tdimc},$ while the RX-side SIM implements $\GR(\phv_{\Rside}) \in \mathbb{C}^{\Rdimc \times \Nr},$ where $\Nt$ and $\Nr$ denote the numbers of TX and RX antenna elements, respectively, and $\phv_{\Tside}$ and $\phv_{\Rside}$ collect the controllable phase shifts of the respective SIM layers\footnote{The antenna dimensions $\Nt$ and $\Nr$ may differ from the latent dimensions $\Tdimc$ and $\Rdimc$, allowing OTA compression at the TX or expansion at the RX; this will be discussed further in Sec.~\ref{sec:sim_schemes}.}. 

The TX and RX communicate over a flat Rayleigh fading MIMO channel \(\mathbf H\in\mathbb C^{\Nr\times\Nt}\), with independent zero-mean circularly symmetric complex Gaussian entries. The received signal is also corrupted by additive white Gaussian noise \(\mathbf v\sim\mathcal{CN}(\mathbf 0,\sigma_v^2\mathbf I)\) appended at the output of the RX-SIM. Assuming unit-power normalized semantic symbols, we parametrize the noise level through the nominal signal-to-noise ratio (SNR), defined as \(\snr\triangleq1/\sigma_v^2\). The resulting baseband model for the received signal of the considered dual-SIM system is:
\begin{equation}
\tilde{\y}
\triangleq
\bsc\,
\GR(\phv_{\Rside})\,
\H\,
\GT(\phv_{\Tside})\,
\x
+
\mathbf{v},
\label{eq:dual_system_io}
\end{equation}
where $\bsc \in \mathbb{C}$ is a scalar calibration factor that adjusts the amplitude and phase of the cascaded transformation to approximate the target semantic alignment operator (see Sec.~\ref{sec:sim_schemes} and~\ref{sec:dual_optim}). The RX then recovers an estimate of the real-valued latent representation as $\hat{\Rlatr} \triangleq \cmap_{\Rside}^{-1}(\tilde{\y})$. The model also encompasses single-SIM configurations as special cases, with the deployment and optimization strategies detailed in Sec.~\ref{sec:sim_schemes}.

\subsection{Dual-SIM Electromagnetic Model}

The considered dual-SIM MIMO SC system comprises an $L_{\Tside}$-layer TX-SIM and an $L_{\Rside}$-layer RX-SIM, collocated with the respective transceivers, as illustrated in Fig.~\ref{fig:system_model}. We describe their electromagnetic transfer matrices below.

\subsubsection{TX-SIM}

The TX latent symbols are embedded into the wave domain by modulating the input SIM layer (layer $0$), which consists of $M^{(\Tside)}_0=\Tdimc$ response-tunable metamaterial elements, as follows:
\begin{align}\label{eq:first_layer}
\Ups^{(\Tside)}_0
\triangleq
\amplconst^{(\Tside)}_0
\mathrm{diag}(\x)
\in \mathbb{C}^{\Tdimc \times \Tdimc},
\end{align}
whose backplate is illuminated by a directive beacon signal~\cite{LML23}, with $\amplconst^{(\Tside)}_0$ denoting an optional amplification factor. Each subsequent layer $\ell=1,\ldots,L_{\Tside}$ comprises $M^{(\Tside)}_\ell$ elements, which are free design parameters for the intermediate layers $\ell<L_{\Tside}$. The last layer is the one radiating into the channel, so it has one element per transmit antenna, i.e., $M^{(\Tside)}_{L_{\Tside}}=\Nt$. Inter-layer propagation follows the geometric-optics model~\cite{AXN23_SIM}. Let $d^{(\Tside)}_{m,m'}$ denote the distance between element $m$ of layer $\ell$ and element $m'$ of layer $\ell-1$, and let $A_{\rm cell}$ and $s_{\rm layer}$ denote the element area and inter-layer spacing, respectively. The propagation matrix $\W^{(\Tside)}_\ell \in\mathbb{C}^{M^{(\Tside)}_\ell \times M^{(\Tside)}_{\ell-1}}$ has entries:
\begin{align}
[\W^{(\Tside)}_\ell]_{m,m'}
&\!\triangleq\!
\frac{s_{\rm layer} A_{\rm cell}}
{(d^{(\Tside)}_{m,m'})^2}
\!\left(
\frac{1}{2\pi d^{(\Tside)}_{m,m'}}
\!-\!\frac{\jmath}{\wavelength}
\right)\!
e^{\jmath 2\pi d^{(\Tside)}_{m,m'}/\wavelength},
\end{align}
where $\wavelength$ is the wavelength. %, all lengths are expressed in the same unit. 
Each element acts as an ideal phase shifter with the following response:
\[
[\resp^{(\Tside)}_\ell]_m
\triangleq
\amplconst^{(\Tside)}_\ell
e^{\jmath\phs^{(\Tside)}_{\ell,m}},
\]
where $\phs^{(\Tside)}_{\ell,m}$ is the tunable phase and $\amplconst^{(\Tside)}_\ell$ is a fixed gain compensating inter-layer attenuation. By defining $\Ups^{(\Tside)}_\ell \triangleq\mathrm{diag}(\resp^{(\Tside)}_\ell)$, the TX-SIM transfer matrix can be expressed as:
\begin{equation}
\GT(\phv_{\Tside})
=
\prod_{\ell=L_{\Tside}}^{1}
\Ups^{(\Tside)}_\ell
\W^{(\Tside)}_\ell
\in\mathbb{C}^{\Nt\times\Tdimc},
\label{eq:GT_dual}
\end{equation}
where matrix products are ordered from right to left, i.e., $\Ups^{(\Tside)}_{L_{\Tside}} \W^{(\Tside)}_{L_{\Tside}} \cdots \Ups^{(\Tside)}_1 \W^{(\Tside)}_1$; this convention is used in the remainder of the paper. Assuming unit-amplitude beacon illumination and $\amplconst^{(\Tside)}_0=1$ in~\eqref{eq:first_layer}, without loss of generality, the signal transmitted through the MIMO channel is given by:
\begin{equation}
\txsig
=
\GT(\phv_{\Tside})
\underbrace{\Ups^{(\Tside)}_0\mathbf{1}}_{=\x}
\in\mathbb{C}^{\Nt},
\label{eq:tx_sim_map}
\end{equation}
where $\mathbf{1}$ is the illumination
vector with $\Tdimc$ ones.

\subsubsection{RX-SIM}

The RX-SIM comprises $L_{\Rside}$ layers. We index its element grids by $\ell=0,\ldots,L_{\Rside}$, where grid $\ell$ has $M^{(\Rside)}_\ell$ elements. Grid $0$ faces the channel, so $M^{(\Rside)}_0=\Nr$; the intermediate grids $\ell=1,\ldots,L_{\Rside}-1$ have free sizes $M^{(\Rside)}_\ell$; and the output grid has $M^{(\Rside)}_{L_{\Rside}}=\Rdimc$ elements. Unlike the TX-SIM, layer $\ell=1,\ldots,L_{\Rside}$ applies phase shifts on grid $\ell-1$ before propagation to grid $\ell$, while the output grid carries no phase shifters. The propagation matrices $\W^{(\Rside)}_\ell \in\mathbb{C}^{M^{(\Rside)}_\ell \times M^{(\Rside)}_{\ell-1}}$ follow the same geometric-optics model with the TX-SIM. The phase-shifter responses are $[\resp^{(\Rside)}_\ell]_m =\amplconst^{(\Rside)}_\ell e^{\jmath\phs^{(\Rside)}_{\ell,m}}$, for $m=1,\ldots,M^{(\Rside)}_{\ell-1}$. With $\Ups^{(\Rside)}_\ell \triangleq\mathrm{diag}(\resp^{(\Rside)}_\ell) \in\mathbb{C}^{M^{(\Rside)}_{\ell-1} \times M^{(\Rside)}_{\ell-1}}$, the RX-SIM transfer matrix becomes as follows:
\begin{equation}
\GR(\phv_{\Rside})
=
\prod_{\ell=1}^{L_{\Rside}}
\W^{(\Rside)}_\ell
\Ups^{(\Rside)}_\ell
\in\mathbb{C}^{\Rdimc\times\Nr}.
\label{eq:GR_dual}
\end{equation}

\subsubsection{Effective Dual-SIM Transformation}

Combining~\eqref{eq:GT_dual} and~\eqref{eq:GR_dual}, the effective electromagnetic transformation is:
\begin{equation}
\Zeff
(\phv_{\Tside},\phv_{\Rside})
\triangleq
\GR(\phv_{\Rside})
\H
\GT(\phv_{\Tside})
\in\mathbb{C}^{\Rdimc\times\Tdimc},
\label{eq:Heff_sys}
\end{equation}
yielding, after substituting in~\eqref{eq:dual_system_io}, the compact input--output relation $\tilde{\y} = \bsc\, \Zeff (\phv_{\Tside},\phv_{\Rside}) \x+\mathbf{v}$.

The trainable parameters are the phase shifts of both SIMs, i.e.:
\[
\phv_{\Tside}
=
\{\{\phs^{(\Tside)}_{\ell,m}\}_{m=1}^{M^{(\Tside)}_\ell}
\}_{\ell=1}^{L_{\Tside}},
\qquad
\phv_{\Rside}
=
\{\{\phs^{(\Rside)}_{\ell,m}\}_{m=1}^{M^{(\Rside)}_{\ell-1}}
\}_{\ell=1}^{L_{\Rside}}.
\]
In the following, we will show that their optimization enables the considered dual-SIM cascade to emulate a semantic alignment operator directly in the wave domain.

\section{Proposed Semantic Alignment Strategies}
\label{sec:optimization}

This section introduces two linear semantic alignment strategies, namely supervised and zero-shot alignment, which serve as target operators for the SIM-based OTA equalization developed in subsequent sections.

\subsection{Supervised Linear Semantic Alignment}

We seek a complex linear operator $\AlignMatL \in \mathbb{C}^{\Rdimc \times \Tdimc}$ that maps TX latent representations into the RX latent space, minimizing the mismatch between the heterogeneous encoders. The operator is learned from a set of semantic pilots (SPs), obtained from a shared subset $\setS\subset\setD$. Their pre-whitened, complex-valued representations are collected in $\X\in\mathbb{C}^{\Tdimc\times|\setS|}$ and $\Y\in\mathbb{C}^{\Rdimc\times|\setS|}$ at the TX and RX, respectively. We are particularly interested in minimizing the Frobenius-norm alignment error, termed \textit{semantic loss} $\SemL$, with $\ell_2$ regularization:
\begin{equation}
\min_{\AlignMatL}\;\;
\underbrace{
\left\lVert \Y-\AlignMatL\X
\right\rVert_F^2
}_{\text{Semantic Loss} \SemL}
+
\reg\left\lVert\AlignMatL\right\rVert_F^2,
\label{eq:linear_prob}
\end{equation}
where $\reg>0$ controls the regularization strength.
The globally optimal solution for this problem is given as:
\begin{equation}
\AlignMatL =
\Y\X^H
\left(
\X\X^H+
\reg\eye{\Tdimc}
\right)^{-1}.
\label{eq:linear_sol}
\end{equation}
The regularization term stabilizes the inversion and controls the magnitude of the operator. Although estimating $\AlignMatL$ requires exchanging paired latent representations, the underlying DNN parameters need not be shared. Once estimated, $\AlignMatL$ provides a fixed linear mapping for aligning incoming TX features with the RX latent space.

\subsection{Zero-Shot Parseval Frame Equalizers}

When semantic pilot exchange is impractical, \textit{Parseval Frame Equalizers} (PFEs) provide a zero-shot alternative~\cite{fiorellino2025frame}. To this end, rather than paired latent representations, PFEs rely on a pre-agreed ordered set of reference anchors $\setA=\{\anch_1,\ldots, \anch_{|\setA|}\}$, processed independently by the TX and RX encoders. Each side constructs its frame operator as:
\begin{equation}
\Fm_{\Tside}
=
\X_{\setA}
\left(
\X_{\setA}^H
\X_{\setA}
\right)^{-1/2},
\,\,
\Fm_{\Rside}
=
\Y_{\setA}
\left(
\Y_{\setA}^H
\Y_{\setA}
\right)^{-1/2},
\label{eq:pfe_def}
\end{equation}
where $\X_{\setA}\in \mathbb{C}^{|\setA|\times\Tdimc}$ and $\Y_{\setA}\in \mathbb{C}^{|\setA|\times\Rdimc}$ collect the respective anchor representations. For full-column-rank anchor matrices, the normalization ensures $\Fm_{\Tside}^H\Fm_{\Tside} =\eye{\Tdimc}$ and $\Fm_{\Rside}^H\Fm_{\Rside}=\eye{\Rdimc}$, yielding Parseval frames in the overcomplete regime. The cross-model semantic alignment operator is then:
\begin{equation}
\AlignMatF
=
\Fm_{\Rside}^H
\Fm_{\Tside}.
\label{eq:pfe_sol}
\end{equation}
Alignment quality depends on anchor selection, as redundant or poorly distributed samples may fail to capture the latent-space geometry. To address this issue, \textit{Prototypical PFEs} (PPFEs)~\cite{fiorellino2025frame} employ \textit{prototypical anchors}, originally introduced in~\cite{huttebraucker2024relative}. These are constructed by clustering the reference samples into $\nclusters\triangleq|\setA|$ clusters and averaging the latent representations of $\nsamples$ randomly selected samples within each cluster, yielding representative anchors that better capture the underlying semantic structure. Since clustering is performed on a single reference model and the resulting sample groups are shared across models, the approach preserves zero-shot alignment without exchanging paired latent representations; further details are provided in~\cite{fiorellino2025frame}.
\section{SIM Design for OTA Semantic Alignment}
\label{sec:sim_schemes}

In this section, we develop a unified and flexible framework for SIM-based OTA semantic alignment, encompassing multiple deployment and optimization strategies. Building on the system model of Section~\ref{sec:system_model} and the semantic operators of the previous Section~\ref{sec:optimization}, our framework accommodates single- and dual-SIM architectures, with independent or joint optimization and different treatments of the MIMO channel $\H$. These alternatives offer different trade-offs in hardware complexity, channel knowledge, and TX--RX coordination. To establish a common formulation for schemes involving individual SIM optimization, let $s \in \{\Tside,\Rside\}$ denote the transceiver side, $\LmatS$ and $\RmatS$ be fixed weighting matrices incorporating channel effects when needed, and let $\TargetS$ be the operator to emulate. The corresponding general optimization problem is:
\begin{equation}
  \min_{\phv_s,\,\bsc_s}
  \bigl\|
  \bsc_s\,\LmatS\,\Gs(\phv_s)\,\RmatS
  -\TargetS
  \bigr\|_F^2.
  \label{eq:generic_loss}
\end{equation}
For fixed phases $\phv_s$, the optimal
complex scaling factor is
\begin{equation}
  \bsc_s
  =
  \frac{\z_s^H\mathbf{t}_s}{\z_s^H\z_s},
  \,\,
  \z_s\triangleq\mathrm{vec}
  \bigl(\LmatS\Gs(\phv_s)\RmatS\bigr),
  \,\,
  \mathbf{t}_s\triangleq\mathrm{vec}(\TargetS).
  \label{eq:beta_single}
\end{equation}
Here, $\Target\in\mathbb{C}^{\Rdimc\times\Tdimc}$ denotes the desired semantic alignment operator, i.e., $\Target=\AlignMatL$ for the supervised solution~\eqref{eq:linear_sol} and $\Target=\AlignMatF$ for the zero-shot solution~\eqref{eq:pfe_sol}, while $\TargetS$ is the corresponding per-SIM target, possibly incorporating channel-dependent factors. Each configuration is specified by the choice of $(\LmatS,\RmatS,\TargetS)$, as summarized in Table~\ref{tab:sim_configs}. The different deployment and optimization strategies are detailed in the following subsections. 

% =======================================================================

\subsection{Mono-SIM}
\label{sec:mono_sim}

The mono-SIM scheme deploys a single SIM at either the TX or RX. We focus on the TX-only configuration, adopted in our preliminary work~\cite{pandolfo2026over} and used here as a baseline. The RX-only variant is considered only in constructive form (i.e., $\bsc_{\Rside}\GR\H\approx\Target$), as reported in Table~\ref{tab:sim_configs}.

\subsubsection{TX-only, equalized}
A digital zero-forcing (ZF) or minimum mean-square error (MMSE) equalizer $\ChannelEq\in\mathbb{C}^{\Nt\times\Nr}$ is applied at the RX to compensate for the MIMO channel. This decouples semantic alignment from channel equalization, allowing the TX-SIM to emulate the target operator independently of $\H$, leading to the following design objective:
\begin{equation}
  \min_{\phv_{\Tside},\,\bsc_{\Tside}}
  \bigl\|
  \bsc_{\Tside}\GT(\phv_{\Tside})
  -\Target
  \bigr\|_F^2.
  \label{eq:mono_eq}
\end{equation}

\subsubsection{TX-only, constructive}
In the constructive approach, the physical channel is exploited as part of the alignment transformation rather than compensated for separately. This requires channel state information (CSI), which can be obtained through channel estimation~\cite{deAlmeida2026}. Specifically, $\H$ is incorporated into the optimization so that the combined channel--SIM response directly approximates the target semantic operator, yielding the design problem:
\begin{equation}
  \min_{\phv_{\Tside},\,\bsc_{\Tside}}
  \bigl\|
  \bsc_{\Tside}\H\GT(\phv_{\Tside})
  -\Target
  \bigr\|_F^2.
  \label{eq:mono_co}
\end{equation}
This removes the need for digital channel
equalization at RX.

% =======================================================================

\begin{table*}[t]
\centering
\caption{SIM-based semantic alignment configurations and their per-SIM training parameters $(\LmatS, \RmatS, \TargetS)$ for problem~\eqref{eq:generic_loss},\\ with $\TargetT,\TargetR$ the SVD parameters of $\Target$ in~\eqref{eq:svd_factor} and $\TargetTtil,\TargetRtil$ the channel-corrected targets in~\eqref{eq:disj_sym_eq_targets}.}
\label{tab:sim_configs}
\setlength{\tabcolsep}{3pt}
\renewcommand{\arraystretch}{1.2}
\begin{tabular}{l ccc ccc c c l}
\multirow{2}{*}{Configuration}
  & \multicolumn{3}{c}{TX SIM}
  & \multicolumn{3}{c}{RX SIM}
  & \multicolumn{3}{c}{Requirements} \\
\cmidrule(lr){2-4}\cmidrule(lr){5-7}\cmidrule(lr){8-10}
  & $\mathbf{B}_{\Tside}$ & $\mathbf{C}_{\Tside}$ & $\TargetT$
  & $\mathbf{B}_{\Rside}$ & $\mathbf{C}_{\Rside}$ & $\TargetR$ & Digital Equalizer & Perfect CSI & Admissible $(\Nt,\Nr)$ \\
\midrule
Mono TX, equalized           & $\mathbf{I}$                                    & $\mathbf{I}$ & $\Target$               & --          & --                                       & --                        & $\checkmark$ & $\checkmark$ & $\Nt=\Rdimc$,\ $\Nr\ge\Nt$ \\
Mono TX, constructive        & $\H$                                    & $\mathbf{I}$ & $\Target$               & --          & --                                       & --                        & -- & $\checkmark$ & $\Nr=\Rdimc$ \\
Mono RX, constructive        & --                                              & --           & --                      & $\mathbf{I}$ & $\H$                              & $\Target$                 & -- & $\checkmark$ & $\Nt=\Tdimc$ \\
\midrule
Disj.-asym. TX, equalized         & $\mathbf{I}$                                    & $\mathbf{I}$ & $\Target$               & $\mathbf{I}$ & $\mathbf{I}$                              & $\PseudoInv$   & -- & $\checkmark$ & $\Nt=\Rdimc$,\ $\Nr\ge\Nt$ \\
Disj.-asym. TX, constructive      & $\mathbf{I}$                                    & $\mathbf{I}$ & $\Target$               & $\mathbf{I}$ & $\H$                              & $\mathbf{I}$               & -- & $\checkmark$ & $\Nt=\Rdimc$,\ $\Nr\ge\Nt$ \\
Disj.-asym. RX, constructive      & $\H$                                    & $\mathbf{I}$ & $\mathbf{I}$               & $\mathbf{I}$ & $\mathbf{I}$                              & $\Target$               & -- & $\checkmark$ & $\Nr=\Tdimc$,\ $\Nt\ge\Nr$ \\
\midrule
Disj.-sym., equalized          & $\mathbf{I}$                                    & $\mathbf{I}$ & $\TargetTtil$ & $\mathbf{I}$ & $\mathbf{I}$                              & $\TargetRtil$ & -- & $\checkmark$ & Any \\
Disj.-sym., constructive       & $\SigHsq\VH^H$    & $\mathbf{I}$ & $\TargetT$             & $\mathbf{I}$ & $\UH\SigHsq$ & $\TargetR$             & -- & $\checkmark$ & Any \\
\midrule
\textbf{Joint, constructive}   & \multicolumn{6}{c}{$\displaystyle\min_{\phv_{\Tside},\phv_{\Rside},\bsc}\;\bigl\|\bsc\,\GR\H\GT - \Target\bigr\|_F^2$ \quad (bilinear, not separable in \eqref{eq:generic_loss})}  & -- & $\checkmark$ & Any \\
\bottomrule
\end{tabular}
\end{table*}

% =======================================================================

\subsection{Disjoint Dual-SIM}
\label{sec:disj_sim}

The disjoint architecture deploys a SIM at both ends of the link, increasing the available wave-domain degrees of freedom while retaining independent optimization. Each SIM solves~\eqref{eq:generic_loss} without requiring the other side's parameters or end-to-end gradients, avoiding the coupling of joint optimization, thus enabling decentralized training. We distinguish \emph{asymmetric} designs, where one SIM emulates the full semantic operator and the other compensates for the channel, from \emph{symmetric} designs, where both side SIMs implement complementary factors of the alignment operator. For the latter, we consider the singular value decomposition (SVD):
\begin{equation}
  \Target = \UT\,\SigT\,\VT^H,
  \label{eq:T_svd}
\end{equation}
where $\UT \in \mathbb{C}^{\Rdimc \times d}$, $\VT \in \mathbb{C}^{\Tdimc \times d}$, $\SigT\triangleq\mathrm{diag}(\sigma_1^E,\ldots,\sigma_d^E)$, and $d\triangleq\mathrm{rank}(\Target)$. This yields the complementary factors
\begin{equation}
  \TargetT=\SigT^{1/2}\VT^H,
  \,\,
  \TargetR=\UT\SigT^{1/2},
  \label{eq:svd_factor}
\end{equation}
satisfying $\TargetR\TargetT=\Target$. On the other hand, channel-aware designs exploit the SVD of the channel matrix $\H=\UH\SigH\VH^H$, where $\UH\in\mathbb{C}^{\Nr\times r}$, $\VH\in\mathbb{C}^{\Nt\times r}$, $\SigH\triangleq \mathrm{diag}(\sigma_1^{(H)},\ldots,\sigma_r^{(H)})$, and $r\triangleq\mathrm{rank}(\H)$. For PPFE alignment, the target operator naturally admits the factorization $\Target=\Fm_{\Rside}^H\Fm_{\Tside}$, yielding $\TargetT=\Fm_{\Tside}$ and $\TargetR=\Fm_{\Rside}^H$ without requiring an additional SVD.

\subsubsection{Asymmetric, equalized}
\label{sec:disj_asym_eq}

In this configuration, the TX-SIM emulates the semantic operator $\Target$ independently of the channel, while the RX-SIM approximates the regularized channel pseudo-inverse $\PseudoInv=\VH \mathrm{diag}\bigl( \sigma_i^{(H)}/((\sigma_i^{(H)})^2+\regeps) \bigr)\UH^H$, with $\regeps>0$. The two SIMs are independently optimized as:
\begin{equation}
  \min_{\phv_{\Tside},\,\bsc_{\Tside}}
  \|\bsc_{\Tside}\GT(\phv_{\Tside})-\Target\|_F^2,
  \,\,
  \min_{\phv_{\Rside},\,\bsc_{\Rside}}
  \|\bsc_{\Rside}\GR(\phv_{\Rside})-\PseudoInv\|_F^2.
  \label{eq:disj_asym_eq}
\end{equation}
Thus, only the RX-SIM requires channel knowledge~\cite{deAlmeida2026}, enabling channel compensation entirely in the wave domain without digital equalization.

\subsubsection{Asymmetric, constructive}
\label{sec:disj_asym_co}

The TX-SIM again emulates $\Target$, while the RX-SIM is optimized jointly with the channel response to approximate an identity mapping, i.e., $\bsc_{\Rside}\GR\H\approx\mathbf{I}$. The corresponding design objectives are expressed as:
\begin{equation}
  \min_{\phv_{\Tside},\,\bsc_{\Tside}}
  \|\bsc_{\Tside}\GT(\phv_{\Tside})-\Target\|_F^2,
  \,\,
  \min_{\phv_{\Rside},\,\bsc_{\Rside}}
  \|\bsc_{\Rside}\GR(\phv_{\Rside})\H-\mathbf{I}\|_F^2.
  \label{eq:disj_asym_co_tx}
\end{equation}
Under exact fitting, the overall cascade satisfies $\bsc\GR^\star\H\GT^\star=\Target$ with $\bsc=\bsc_{\Tside}\bsc_{\Rside}$. Unlike pseudo-inverse fitting, the channel enters the loss directly, weighting errors according to its singular values and reducing the penalty along weak channel directions. The RX-oriented counterpart exchanges the SIM roles and follows the same constructive principle (see Table~\ref{tab:sim_configs}).

\subsubsection{Symmetric, equalized}
\label{sec:disj_sym_eq}

In the symmetric equalized scheme, the semantic operator and channel compensation are distributed across both end SIMs. Specifically, the channel SVD is incorporated into the individual targets:
\begin{equation}
  \TargetTtil=\VH\SigHisq\TargetT,
  \,\,
  \TargetRtil=\TargetR\SigHisq\UH^H,
  \label{eq:disj_sym_eq_targets}
\end{equation}
where $\SigHisq\triangleq\mathrm{diag}\Bigl(
\sqrt{\frac{\sigma_i^{(H)}}{(\sigma_i^{(H)})^2+\regeps}}
\Bigr)$.
Both end SIMs are then independently optimized as:
\begin{equation}
  \min_{\phv_{\Tside},\,\bsc_{\Tside}}
  \|\bsc_{\Tside}\GT(\phv_{\Tside})-\TargetTtil\|_F^2,
  \,\,
  \min_{\phv_{\Rside},\,\bsc_{\Rside}}
  \|\bsc_{\Rside}\GR(\phv_{\Rside})-\TargetRtil\|_F^2.
  \label{eq:disj_sym_eq}
\end{equation}
Unlike constructive designs, channel information is embedded in the target matrices rather than the fitting objectives. Under exact SIM fitting, the resulting cascade satisfies
\begin{align}
  \bsc\GR^\star\H\GT^\star
  &=
  \TargetR\SigHisq
  \SigH\SigHisq\TargetT\approx\Target,
  \label{eq:disj_sym_co_cancel}
\end{align}
where the approximation accounts for the regularization-induced residual distortion.

\subsubsection{Symmetric, constructive}
\label{sec:disj_sym_co}

Unlike the equalized scheme, the constructive variant incorporates channel-dependent weighting directly into the per-SIM optimization objectives:
\begin{align}
  \min_{\phv_{\Tside},\,\bsc_{\Tside}}
  &\|\bsc_{\Tside}\SigHsq\VH^H
  \GT(\phv_{\Tside})-\TargetT\|_F^2,
  \notag\\
  \min_{\phv_{\Rside},\,\bsc_{\Rside}}
  &\|\bsc_{\Rside}\GR(\phv_{\Rside})
  \UH\SigHsq-\TargetR\|_F^2,
  \label{eq:disj_sym_co}
\end{align}
where $\SigHsq=\mathrm{diag}\bigl(
\sigma_i^{(H)}/\sqrt{\sigma_i^{(H)}+\regeps} \bigr)$. This formulation exploits the channel singular modes during training while retaining independent optimization of the SIM at both ends.

% % =======================================================================

\subsection{Joint Dual-SIM}
\label{sec:joint_sim}

The joint constructive scheme optimizes both end SIMs over the complete TX--channel--RX cascade, directly incorporating the physical channel into semantic alignment. Unlike the disjoint schemes, the SIM parameters are coupled through $\H$, leading to the following non-convex design objective:
\begin{equation}
  \min_{\phv_{\Tside},\,\phv_{\Rside},\,\bsc}
  \mathcal{L}_{\mathrm{J}}
  (\phv_{\Tside},\phv_{\Rside},\bsc)
  \triangleq
  \bigl\|
  \bsc\GR(\phv_{\Rside})
  \H\GT(\phv_{\Tside})-\Target
  \bigr\|_F^2.
  \label{eq:joint_obj_sec}
\end{equation}
For fixed SIM phases, the optimal scaling is $\bsc^\star= \langle\Zeff,\Target\rangle_F/\|\Zeff\|_F^2$, where $\Zeff=\GR\H\GT$. The phase configurations are jointly designed through the alternating projected gradient algorithm developed in Section~\ref{sec:dual_optim}, eliminating the need for digital channel equalization. By optimizing the entire cascade, the joint scheme can exploit interactions between the two SIM and the channel that are not explicitly captured by independent per-SIM fitting. This increased design flexibility comes at the cost of global channel knowledge~\cite{deAlmeida2026}, TX--RX coordination, and a coupled non-convex optimization problem.

Table~\ref{tab:sim_configs} summarizes the configurations and their requirements. The mono-SIM baseline offers the simplest hardware deployment, and disjoint dual-SIM schemes enable independent optimization, whereas the joint scheme instead optimizes the full cascade and is therefore the focus of Sec.~\ref{sec:dual_optim}.

\textbf{Remark 1 (Antenna dimensions and spatial compression).}
In mono-SIM and asymmetric designs, the dimensions of the emulated operators impose constraints on $\Nt$ and $\Nr$ (see Table~\ref{tab:sim_configs}). In contrast, symmetric and joint schemes allow the channel to act as a bottleneck, enabling OTA compression and expansion of latent representations even when the antenna dimensions are smaller than the corresponding latent spaces. The rank of the effective operator is bounded by $\min(\Nt,\Nr,\mathrm{rank}(\H))$, limiting the achievable compression.

\textbf{Remark 2 (Information exchange requirements).}
For PPFE-based symmetric designs, each side constructs its own target factor from local anchor representations, with $\TargetT=\Fm_{\Tside}$ and $\TargetR=\Fm_{\Rside}^H$, requiring only a pre-agreed anchor set and ordering. Channel-dependent factors are obtained from the corresponding channel singular vectors and values, assuming a consistent SVD convention. Conversely, schemes emulating the full semantic operator require exchanging the relevant PPFE factor, while supervised alignment relies on semantic pilots to estimate $\Target$ and its factorization.

\section{Dual-SIM Optimization\texorpdfstring{\\}{ } for the Joint Constructive Scheme}
\label{sec:dual_optim}

Among the configurations surveyed in Section~\ref{sec:sim_schemes}, we focus on the \emph{joint constructive} scheme, which provides the richest emulation capability at the cost of solving a bilinear optimization problem over the phase parameters of both end SIMs. We design $(\GT,\GR)$ by minimizing the joint emulation loss $\LossJ(\phv_{\Tside},\phv_{\Rside},\bsc)$ in~\eqref{eq:joint_obj_sec} under the phase constraints of the SIMs:
\begin{subequations}
\label{eq:dual_sim_opt}
\begin{align}
\min_{\phv_{\Tside},\phv_{\Rside},\bsc}\quad
& \LossJ(\phv_{\Tside},\phv_{\Rside},\bsc)
\label{eq:dual_sim_opt_a}
\\
\mathrm{s.t.}\quad
& \GT(\phv_{\Tside}) \ \text{as in} \ \eqref{eq:GT_dual}, \,\,
\GR(\phv_{\Rside}) \ \text{as in} \ \eqref{eq:GR_dual},
\label{eq:dual_sim_opt_b}
\\
& \phs^{(\Tside)}_{\ell,m}\in[0, 2\pi)\, \forall \ell,m,
\,\,
\phs^{(\Rside)}_{\ell,m}\in[0, 2\pi)\, \forall \ell,m.
\label{eq:dual_sim_opt_c}
\end{align}
\end{subequations}
This problem is non-convex due to the bilinear coupling between $\GT$ and $\GR$ in the objective and the constant-modulus constraints in~\eqref{eq:dual_sim_opt_c}. We address it with two complementary solvers: an alternating projected gradient algorithm with convergence guarantees (Sec.~\ref{sec:alt_pg}), and a deep-unrolled network that reduces the cost of reconfiguring the SIMs when the channel changes (Sec.~\ref{sec:deep_unrolling}).

\subsection{Alternating Projected Gradient Algorithm}
\label{sec:alt_pg}

We employ an alternating minimization strategy, where $(\phv_{\Tside},\phv_{\Rside})$ are updated by gradient-based steps while $\bsc$ admits a closed-form update.

\textbf{Closed-form update for $\bsc$:}
Given $\phv_{\Tside}$ and $\phv_{\Rside}$ at iteration $k$, let us define:
\begin{equation}
\Zeff^{k} \triangleq \GR\left(\phv_{\Rside}^{k}\right)\H\GT\left(\phv_{\Tside}^{k}\right).
\label{eq:Z_def}
\end{equation}
Assuming $\|\Zeff^{k}\|_F>0$, the optimal $\bsc^{k}$ minimizing $\| \bsc \Zeff^{k} - \Target\|_F^2$ is given by:
\begin{equation}
\bsc^{k} =
\frac{\langle \Zeff^{k}, \Target\rangle}{\|\Zeff^{k}\|_F^2}
=
\frac{\mathrm{vec}(\Zeff^{k})^H\,\mathrm{vec}(\Target)}
{\mathrm{vec}(\Zeff^{k})^H\,\mathrm{vec}(\Zeff^{k})},
\label{eq:beta_closed_form_dual}
\end{equation}
where $\langle \mathbf{M},\mathbf{N}\rangle \triangleq \mathrm{tr}(\mathbf{M}^H \mathbf{N})$ denotes the Frobenius inner product. This least-squares scaling is the one used in~\cite{An2024SIM2DDFT_ICC} to fit a single SIM to a target matrix, and the same closed-form expression also appears in~\cite{AXN23_SIM}.

\textbf{Projected gradient updates for $(\phv_{\Tside},\phv_{\Rside})$:}
For fixed $\bsc^k$, the SIM phase vectors are updated sequentially through the following projected gradient steps:
\begin{align}
\phv_{\Tside}^{k+1}
&=
\Pi_{[0,2\pi]}\!\Big(
\phv_{\Tside}^{k}
-
\lr_{\Tside}
\nabla_{\phv_{\Tside}}
\LossJ
\big(
\phv_{\Tside}^{k},
\phv_{\Rside}^{k},
\bsc^{k}
\big)
\Big),
\label{eq:xiT_update}
\\
\phv_{\Rside}^{k+1}
&=
\Pi_{[0,2\pi]}\!\Big(
\phv_{\Rside}^{k}
-
\lr_{\Rside}
\nabla_{\phv_{\Rside}}
\LossJ
\big(
\phv_{\Tside}^{k+1},
\phv_{\Rside}^{k},
\bsc^{k}
\big)
\Big),
\label{eq:xiR_update}
\end{align}
where $\lr_{\Tside},\lr_{\Rside}>0$ are the respective step sizes.
The RX update uses the newly computed TX phases, while $\bsc^k$ remains fixed throughout both updates. The gradients in \eqref{eq:xiT_update} and \eqref{eq:xiR_update} are obtained by differentiating the SIM transfer matrices \eqref{eq:GT_dual}--\eqref{eq:GR_dual} and the effective cascade~\eqref{eq:Z_def}.

Algorithm~\ref{alg:dual_sim} summarizes the proposed alternating optimization procedure for jointly configuring the TX and RX SIM to emulate the semantic operator $\Target$, by exploiting the structure of the channel $\H$ within the end-to-end cascade $\bsc\,\GR\H\GT$.
\begin{algorithm}[t]
\caption{Projected Gradient Dual-SIM Optimizer}
\label{alg:dual_sim}
\begin{algorithmic}[1]
\Require Target operator
$\Target\in\mathbb{C}^{\Rdimc\times\Tdimc}$,
channel $\H\in\mathbb{C}^{\Nr\times\Nt}$,
SIM propagation matrices
$\{\W^{(\Tside)}_\ell\}_{\ell=1}^{L_{\Tside}}$
and
$\{\W^{(\Rside)}_\ell\}_{\ell=1}^{L_{\Rside}}$,
step sizes $\lr_{\Tside},\lr_{\Rside}$,
maximum iterations $\maxIter$.
\Ensure Phase configurations
$\phv_{\Tside}^\star,\phv_{\Rside}^\star$
and scaling $\bsc^\star$.

\State Initialize
$\phv_{\Tside}^{0},\phv_{\Rside}^{0}$
in $[0,2\pi)$.
\For{$k=0,1,\ldots,\maxIter-1$}
    \State Compute
    $\Zeff^{k}\leftarrow
    \GR(\phv_{\Rside}^{k})\H
    \GT(\phv_{\Tside}^{k})$.
    \State Update
    $\bsc^{k}\leftarrow
    \langle\Zeff^{k},\Target\rangle/
    \|\Zeff^{k}\|_F^2$
    using~\eqref{eq:beta_closed_form_dual}.
    \State Update TX phases
    using~\eqref{eq:xiT_update}.
    \State Update RX phases
    using~\eqref{eq:xiR_update}.
\EndFor
\State Set
$\phv_{\Tside}^\star
=\phv_{\Tside}^{\maxIter}$
and
$\phv_{\Rside}^\star
=\phv_{\Rside}^{\maxIter}$.
\State Compute
$\Zeff^\star\leftarrow
\GR(\phv_{\Rside}^\star)\H
\GT(\phv_{\Tside}^\star)$.
\State Set
$\bsc^\star\leftarrow
\langle\Zeff^\star,\Target\rangle/
\|\Zeff^\star\|_F^2$.
\end{algorithmic}
\end{algorithm}

\textbf{Convergence Properties:}
We now establish the convergence properties of Algorithm~\ref{alg:dual_sim}. Specifically, we express the feasible phase set in the compact form $\phset\triangleq\phset_{\Tside}\times\phset_{\Rside}$, where $\phset_{\Tside}\triangleq[0,2\pi]^{\nph_{\Tside}}$ and $\phset_{\Rside}\triangleq[0,2\pi]^{\nph_{\Rside}}$, with $\nph_{\Tside}\triangleq\sum_{\ell=1}^{L_{\Tside}} M_\ell^{(\Tside)}$ and $\nph_{\Rside}\triangleq\sum_{\ell=1}^{L_{\Rside}} M_{\ell-1}^{(\Rside)}$ denoting the total number of tunable phase shifts at the TX and RX, respectively. We consider closed phase intervals to ensure well-defined Euclidean projections; their endpoints are physically equivalent due to the $2\pi$-periodicity of the SIM responses. Let us also define the reduced objective:
\begin{equation}
\bar F(\phv_{\Tside},\phv_{\Rside})
\triangleq
\min_{\bsc\in\mathbb{C}}
\LossJ(\phv_{\Tside},\phv_{\Rside},\bsc),
\label{eq:reduced_obj}
\end{equation}
whose minimizer is given by \eqref{eq:beta_closed_form_dual}. The following theorem shows that, under Lipschitz block gradients and sufficiently small step sizes, Algorithm~\ref{alg:dual_sim} yields non-increasing objective values and first-order stationary limit points.

\begin{theorem}
\label{thm:conv_dualsim}
Assume that the effective cascade satisfies
$z_{\min}\triangleq
\min_{\phv\in\phset}\|\Zeff\|_F>0$,
and define
$\bsc_{\max}\triangleq
\|\Target\|_F/z_{\min}$.
Let $\lip_{\Tside},\lip_{\Rside}>0$
be Lipschitz constants of the corresponding
block gradients of $F$, uniform over the
other phase block and $|\bsc|\le\bsc_{\max}$. Let $\{(\phv_{\Tside}^{k},\phv_{\Rside}^{k},
\bsc^{k})\}_{k\ge0}$ be the sequence generated
by Algorithm~\ref{alg:dual_sim}, with step sizes
$0<\lr_{\Tside}<1/\lip_{\Tside}$ and
$0<\lr_{\Rside}<1/\lip_{\Rside}$. Then:
\begin{enumerate}
\item The sequence
$\{\bar F(\phv_{\Tside}^{k},
\phv_{\Rside}^{k})\}$
is non-increasing and convergent.

\item Every limit point
$\phv^\star=
(\phv_{\Tside}^\star,\phv_{\Rside}^\star)$
satisfies the first-order stationarity
conditions of the closed-box formulation
of~\eqref{eq:dual_sim_opt}, namely:
\[
\nabla_{\phv_s}\bar F(\phv^\star)^\top
(\phv_s-\phv_s^\star)\ge0,
\quad
\forall\phv_s\in\phset_s,
\quad s\in\{\Tside,\Rside\}.
\]
The associated $\bsc^\star$ is the
exact minimizer given by
\eqref{eq:beta_closed_form_dual}.
\end{enumerate}
\end{theorem}

\begin{proof}
See Appendix~\ref{app:conv_proof}.
\end{proof}

\subsection[Deep Unrolling for Dual-SIM Optimization]
{Deep Unrolling for Dual-SIM Optimization}
\label{sec:deep_unrolling}

The alternating projected gradient algorithm of Algorithm~\ref{alg:dual_sim} guarantees monotone descent, but may require many iterations due to the non-convex coupling between the two end SIMs and the large number of phase variables. Moreover, since all the designs rely on perfect CSI, any channel variation requires solving the optimization problem anew. To avoid this, we present a model-based deep-unrolled architecture~\cite{Shlezinger2023MBLearning,Balatsoukas2019DeepUnfolding} that is trained offline on channel realizations drawn from a known distribution, and adapts to the instantaneous channel with only $K$ gradient-based updates with learnable per-coordinate step sizes and momentum. Unlike the alternating algorithm, the proposed network updates the TX and RX phases simultaneously using normalized gradient signals. To enforce the phase constraints without explicit projection, we parametrize the phases through unconstrained variables $\logit_{\Tside}\in\mathbb{R}^{\nph_{\Tside}}$ and $\logit_{\Rside}\in\mathbb{R}^{\nph_{\Rside}}$, as $\phv_s=2\pi\sigm(\logit_s)$, $s\in\{\Tside,\Rside\}$, where $\sigm(\cdot)$ denotes the element-wise sigmoid function. This parametrization satisfies the constraints in~\eqref{eq:dual_sim_opt_c} by construction. Let $\logit_s^k$ denote the unconstrained phase variables at unrolled layer $k$, and define the effective cascade $\Zeff(\logit_{\Tside},\logit_{\Rside}) \triangleq \GR(\logit_{\Rside}) \H \GT(\logit_{\Tside})$. The normalized emulation error is:
\begin{equation}
\nmse(\logit_{\Tside},\logit_{\Rside})
\triangleq
\min_{\bsc\in\mathbb{C}}
\frac{
\|\bsc\Zeff(\logit_{\Tside},\logit_{\Rside})
-\Target\|_F^2
}{
\|\Target\|_F^2
},
\label{eq:nmse_def}
\end{equation}
solved by $\bsc$ given in~\eqref{eq:beta_closed_form_dual}. At each unrolled layer, the descent direction of each SIM is the negative gradient of the emulation error, normalized by the root-mean-square (RMS):
\begin{equation}
\innov_s^k
\triangleq
-\rmsnorm\left(
\nabla_{\logit_s}
\nmse(\logit_{\Tside}^k,\logit_{\Rside}^k)
\right),
\quad s\in\{\Tside,\Rside\}.
\label{eq:gT_unrolled}
\end{equation}
The operator $\rmsnorm(\cdot)$ normalizes the gradient to unit RMS magnitude over all phase variables of the corresponding SIM, separately for the TX and RX and for each channel realization.

\textbf{Diagonal updates with momentum:}
To increase the flexibility of the unrolled updates, we replace the scalar step sizes of Algorithm~\ref{alg:dual_sim} with learnable per-coordinate operators. Unlike full matrix-valued operators, which would introduce a quadratic number of trainable parameters, this diagonal parametrization assigns three learnable quantities to each phase variable at every unrolled layer: a momentum coefficient, a step size, and a scaling factor. For each SIM $s\in\{\Tside,\Rside\}$, the momentum coefficients are parametrized as $\mom_s^k=\sigm(\hat{\mom}_s^k) \in(0,1)^{\nph_s}$, where $\hat{\mom}_s^k\in\mathbb{R}^{\nph_s}$ is a trainable vector. These coefficients govern an exponential moving average (EMA) of the descent directions:
\begin{align}
\mombuf_s^k
&=\mom_s^k\odot\mombuf_s^{k-1}
+(\mathbf{1}-\mom_s^k)\odot\innov_s^k,
\label{eq:ema_v}\\
\hat{\innov}_s^k
&=\mombuf_s^k\oslash
\left(\mathbf{1}-\prod_{j=0}^{k}\mom_s^j\right),
\label{eq:ema_bcorr}
\end{align}
where $\mombuf_s^{-1}=\mathbf{0}$, $\mombuf_s^k$ is the recursively updated momentum state, and the product is element-wise. Equation~\eqref{eq:ema_bcorr} corrects the initialization bias of the EMA, following the principle used in Adam~\cite{Kingma2015Adam}. The TX and RX logits are then updated simultaneously from the same iterate:
\begin{equation}
\logit_s^{k+1}
=\scal_s^k\odot\logit_s^k
+\stepw_s^k\odot\hat{\innov}_s^k,
\qquad s\in\{\Tside,\Rside\},
\label{eq:diag_tx}
\end{equation}
where $\scal_s^k,\stepw_s^k \in\mathbb{R}^{\nph_s}$ are trainable scaling and step-size vectors, respectively. Without momentum and with $\scal_s^k=\mathbf{1}$, \eqref{eq:diag_tx} reduces to a normalized gradient step with learned per-coordinate step sizes. Collecting the trainable parameters gives $$\params=\{\hat{\mom}_s^k,\stepw_s^k, \scal_s^k\}_{s\in\{\Tside,\Rside\},\,k=0}^{K-1}.$$

With $n=\nph_{\Tside}+\nph_{\Rside}$, the architecture requires $3Kn$ trainable parameters, compared with $\mathcal{O}(Kn^2)$ for dense update operators. The parameters $\params$ are optimized end-to-end using Adam~\cite{Kingma2015Adam}, by backpropagating through the $K$ unrolled layers and the SIM forward models.

\textbf{Warm start:}
Before applying the unrolled network, we obtain a common initialization for the TX and RX phases by running $\wsSteps$ cycles of alternating optimization on the joint emulation loss $\LossJ(\phv_{\Tside},\phv_{\Rside},\bsc)$ with $\H=\mathbf{I}$. This semantic-only optimization yields the initial logits $(\logit_{\Tside}^0,\logit_{\Rside}^0)$, which are computed once and kept fixed across channel realizations. The warm start thus provides an informed, reproducible initialization, leaving channel-dependent adaptation to the subsequent unrolled updates.

\textbf{Unrolled forward pass:}
Algorithm~\ref{alg:unrolled} summarizes the inference procedure for a given channel $\H$. The learned parameters $\params$ are shared across channel realizations, whereas the logits and momentum states evolve independently for each realization. Starting from the common warm start and zero momentum states, each layer evaluates the normalized emulation error \eqref{eq:nmse_def}, computes both descent directions through~\eqref{eq:gT_unrolled}, and applies the momentum and logit updates \eqref{eq:ema_v}--\eqref{eq:diag_tx}. The TX and RX descent directions are evaluated at the same iterate, so both end SIMs are updated simultaneously. Each layer requires one evaluation of the effective cascade and one reverse-mode automatic differentiation pass to obtain both gradients. Consequently, inference through $K$ layers requires $K$ gradient evaluations, without explicitly forming Jacobian matrices or performing matrix inversions. The complex scaling $\bsc$ is not maintained as a network state; it is computed in closed form from the final effective cascade using~\eqref{eq:beta_closed_form_dual}.

\begin{algorithm}[t]
\caption{Deep-Unrolled Dual-SIM Optimizer}
\label{alg:unrolled}
\begin{algorithmic}[1]
\Require Target $\Target$, channel $\H$,
learned parameters $\params$, depth $K$,
warm-start logits
$(\logit_{\Tside}^0,\logit_{\Rside}^0)$.
\Ensure Phases
$\phv_{\Tside}^\star,\phv_{\Rside}^\star$
and scaling $\bsc^\star$.
\State Initialize
$\mombuf_{\Tside}^{-1}\leftarrow\mathbf{0}$,
$\mombuf_{\Rside}^{-1}\leftarrow\mathbf{0}$.
\For{$k=0,\ldots,K-1$}
    \State Evaluate $\nmse$ and its logit
    gradients at
    $(\logit_{\Tside}^k,\logit_{\Rside}^k)$
    
    \hspace{-0.163cm}using~\eqref{eq:nmse_def}.
    \State Compute $\innov_s^k$ for both end SIMs
    using~\eqref{eq:gT_unrolled}.
    \State Update $\mombuf_s^k$ and
    $\hat{\innov}_s^k$ using
    \eqref{eq:ema_v} and \eqref{eq:ema_bcorr}.
    \State Update both $\logit_s^{k+1}$
    simultaneously using~\eqref{eq:diag_tx}.
\EndFor
\State Set
$\phv_s^\star\leftarrow
2\pi\sigm(\logit_s^K)$
for $s\in\{\Tside,\Rside\}$.
\State Compute
$\Zeff^\star\leftarrow
\GR(\phv_{\Rside}^\star)
\H\GT(\phv_{\Tside}^\star)$.
\State Compute $\bsc^\star$
using~\eqref{eq:beta_closed_form_dual}.
\end{algorithmic}
\end{algorithm}

\textbf{Training with deep supervision:}
We train the shared parameters $\params$ over a dataset of channel realizations $\{\H_i\}_{i=1}^{N_{\mathrm{tr}}}$, without requiring reference phase solutions. For channel $\H_i$, the effective cascade after layer $k$ is for $\phv_{s,i}^k=2\pi\sigm(\logit_{s,i}^k)$:
\begin{equation}
\Zeff_i^k
=
\GR(\phv_{\Rside,i}^k)
\H_i
\GT(\phv_{\Tside,i}^k).
\label{eq:ZK_unrolled}
\end{equation}
The corresponding normalized emulation error $\nmse_i^k$ is computed using~\eqref{eq:nmse_def}, with the optimal complex scaling implicitly evaluated as in \eqref{eq:beta_closed_form_dual}. To supervise intermediate as well as final outputs, we minimize the weighted average of the errors across all layers:
\begin{equation}
\LossTrain(\params)
=
\frac{1}{N_{\mathrm{tr}}}
\sum_{i=1}^{N_{\mathrm{tr}}}
\frac{\sum_{k=1}^{K}
\dsw_k\nmse_i^k}
{\sum_{k=1}^{K}\dsw_k},
\,\,
\dsw_k=\dsratio^{\frac{K-k}{K-1}}.
\label{eq:train_loss_deep_sup}
\end{equation}
For $K>1$, the parameter $\dsratio\in(0,1]$ determines the ratio between the first- and last-layer weights. Setting $\dsratio=1$ assigns equal weights, whereas $\dsratio<1$ progressively emphasizes later layers. For $K=1$, we set $\dsw_1=1$. This deep-supervision strategy provides direct training signals to intermediate outputs, facilitating the optimization of deeper unrolled networks. The parameters $\params$ are trained using Adam~\cite{Kingma2015Adam} by backpropagating \eqref{eq:train_loss_deep_sup} through the unrolled layers. To avoid second-order derivatives, the descent directions \eqref{eq:gT_unrolled}, including their RMS normalization, are treated as constants during backpropagation. Gradients thus propagate through the momentum and logit updates \eqref{eq:ema_v}--\eqref{eq:diag_tx}, but not through the descent directions.

\textbf{Progressive training:}
To train architectures of increasing depth, consider $K_{\mathrm{list}}\triangleq[k_0<k_1<\cdots<k_M]$. Under independent training, each architecture is initialized and optimized separately. Instead, our progressive strategy transfers the parameters learned by a shallower network to the corresponding layers of the next, deeper network. Specifically, after training the model of depth $k_i$, we initialize the first $k_i$ layers of the model of depth $k_{i+1}$ as follows:
\begin{equation}
\params^{(k_{i+1})}\big|_{k<k_i}
\leftarrow
\params^{(k_i)},
\label{eq:progressive_init}
\end{equation}
while the additional layers $k=k_i,\ldots,k_{i+1}-1$ receive their default initialization. Since all models use the same warm-start logits, the first $k_i$ layers of the deeper network initially reproduce the output of the trained shallower model. The additional layers can then refine this solution, while subsequent training jointly adjusts the entire network. Progressive initialization thereby reuses previously learned update schedules instead of learning each depth from scratch.

\textbf{Supervised extension:}
The proposed training procedure relies solely on the emulation loss and does not require reference phase configurations. An alternative, not considered in our experiments, is imitation learning from solutions generated offline by a high-accuracy optimizer, such as Algorithm~\ref{alg:dual_sim} with a large iteration budget. Given reference solutions $\{(\Target_i,\H_i, \phv_{\Tside,i}^\star, \phv_{\Rside,i}^\star)\}_{i=1}^{N_{\mathrm{tr}}}$, one could minimize the following phase-regression loss:
\begin{equation}
\LossSup(\params)
\triangleq
\frac{1}{N_{\mathrm{tr}}}
\sum_{i=1}^{N_{\mathrm{tr}}}
\sum_{s\in\{\Tside,\Rside\}}
\left\|
\mathrm{wrap}\left(
\phv_{s,i}^{K}
-\phv_{s,i}^\star
\right)
\right\|_2^2,
\label{eq:phase_supervised_loss}
\end{equation}
where $\mathrm{wrap}(\cdot)$ computes phase differences modulo $2\pi$ element-wise, e.g., $\mathrm{wrap}(x)= \operatorname{atan2}(\sin x,\cos x)$.
\begin{figure*}[!t]
  \centering
  \includegraphics[width=0.99\linewidth]{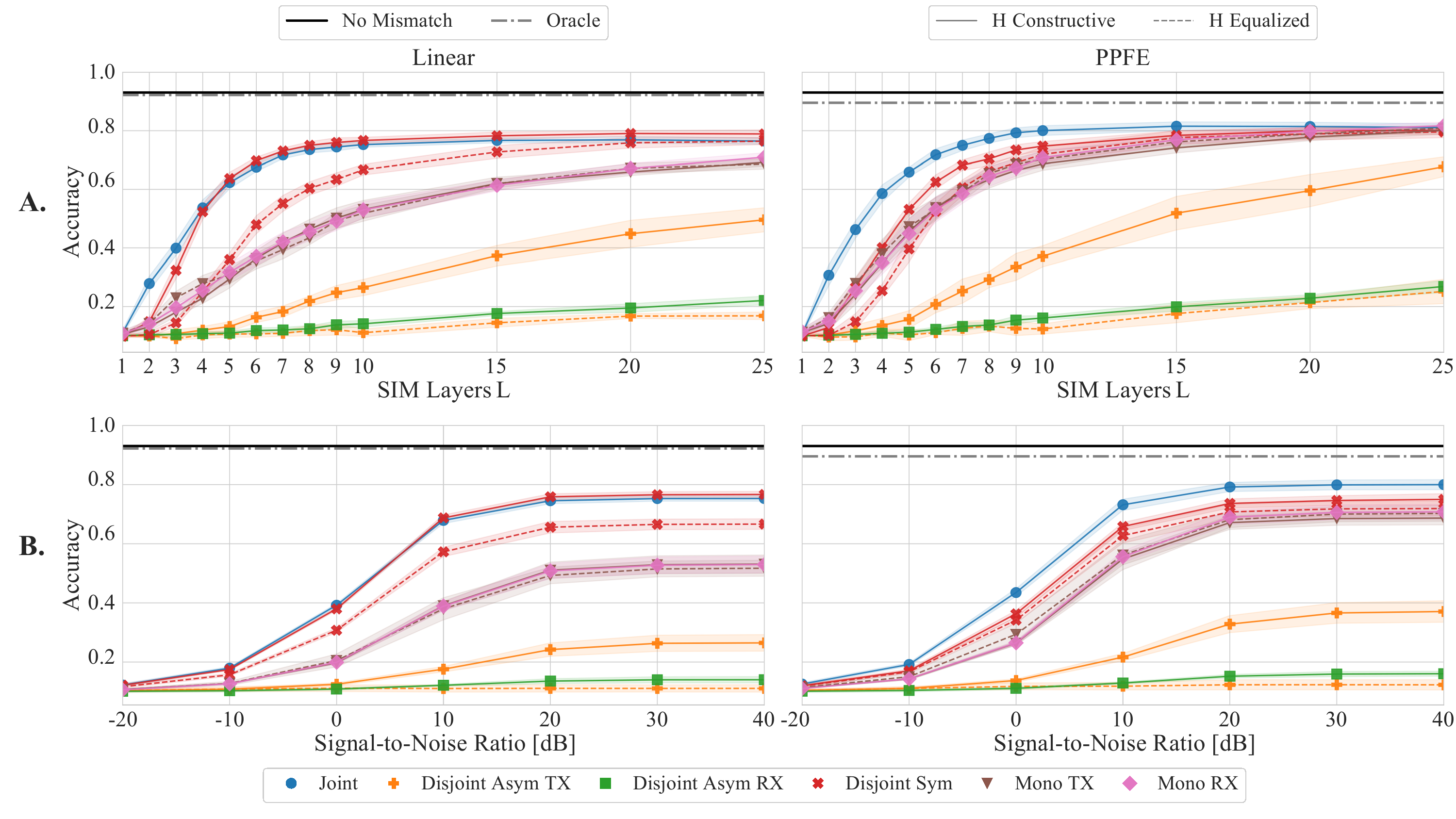}
  \caption{Accuracy for the Linear (left) and PPFE (right) targets with $M_\ell=32\!\times\!32$ $\ell=1,\ldots,L$. \textbf{Row A:} versus the layers per SIM $L$ at infinite SNR. \textbf{Row B:} versus SNR$_{[\mathrm{dB}]}$ with $L=10$. Channel: $20\!\times\!20$ (joint, disjoint symmetric), $192\!\times\!192$ (mono TX, disjoint asymmetric TX), $384\!\times\!384$ (mono RX, disjoint asymmetric RX). Bands: $95\%$ confidence intervals (CI) over six seeds.}
  \label{fig:layers_snr}
\end{figure*}

\section{Numerical Results and Discussion}
\label{sec:numerical}

In this section, we evaluate the performance\footnote{The source code for the paper's numerical evaluation is available at: \href{https://github.com/SPAICOM/semantic-alignment-via-dual-sim.git}{https://github.com/SPAICOM/semantic-alignment-via-dual-sim.git.}} of the proposed dual-SIM semantic alignment framework through two experiments:
\textbf{Experiment~A} (Section~\ref{sec:exp_arch}) benchmarks the emulation capabilities and alignment performance of all schemes of Table~\ref{tab:sim_configs} versus the number of SIM layers and SNR; and 
\textbf{Experiment~B} (Section~\ref{sec:exp_unrolling}) evaluates the deep-unrolled optimizer presented in Section~\ref{sec:deep_unrolling}, in terms of convergence speed and generalization to unseen channel realizations drawn from a known distribution.

%\noindent 
\textbf{General setup.} Experiments A and B used the \textsc{CIFAR-10}~\cite{krizhevsky2009learning} split of the \textsc{Semasia} benchmark~\cite{pandolfo2026semasia}, which provides embeddings pre-computed by pretrained \textit{timm}~\cite{rw2019timm} models: ViT-Base/16~\cite{Dosovitskiy2021ViT} at the TX ($\Tdim=768$ real features, $\Tdimc=384$ complex symbols after $\cmap_{\Tside}$) and ViT-Small/16 at the RX ($\Rdim=384$, $\Rdimc=192$ after $\cmap_{\Rside}$). The alignment quality was measured by a two-layer multilayer perceptron (MLP) classifier (one hidden layer of $512$ Tanh units) playing the role of the pre-trained RX decoder. This classifier was trained for each encoder pair and seed on the \emph{original} RX embeddings (Adam, learning rate $10^{-3}$, early stopping), kept fixed, and fed at test time with the aligned embeddings $\hat{\mathbf{s}}_{\Rside}$, so that any accuracy loss is due to residual misalignment and noise.

\subsection{Emulation Capabilities and Alignment Performance}
\label{sec:exp_arch}

%\noindent 
\textbf{Experiment-specific setup.}
The channel was a flat Rayleigh fading matrix $\H\in\mathbb{C}^{\Nr\times\Nt}$ with i.i.d.\ $\mathcal{CN}(0,1)$ entries, whose size depended on the scheme (Fig.~\ref{fig:layers_snr}, Table~\ref{tab:sim_configs}). For the joint and disjoint symmetric schemes the $20\times20$ cascade matrix $\GR\H\GT$ compressed the $\Tdimc=384$ TX symbols onto $\Nt=20$ transmit antennas and expanded them to the $\Rdimc=192$ RX symbols. The disjoint symmetric target factors had inner dimension $20$ (Section~\ref{sec:disj_sim}), and the other schemes emulated the full operator $\Target$. Here, $L$ is the number of layers \emph{per SIM}. Each layer had $M_\ell=32\times32$ meta-atoms with cell width $\wavelength/2$, $\wavelength=5$\,mm, the inter-layer spacing was $s_{\rm layer}=5\wavelength$, unless stated otherwise, and no amplification was used (i.e., $\amplconst_l=1$ $\forall \ell$). The phases were optimized on unconstrained logits $\logit$, with $\phv=2\pi\sigm(\logit)$, initialized at zero (all phases equal to $\pi$), and the phases with the lowest training loss were retained. The joint scheme alternated between the SIM at the two ends, as in Section~\ref{sec:dual_optim}, with $150$ cycles of $100$ Adam steps~\cite{Kingma2015Adam} per SIM and $\bsc$ recomputed in closed form~\eqref{eq:beta_closed_form_dual} before every step; the mono and disjoint schemes trained each SIM independently for $3000$ Adam steps with learning rate $0.1$. All schemes were run until convergence of the training loss, with learning rates selected by grid search (joint scheme: $0.03$ at the TX and $0.07$ at the RX, $0.04$ for Linear). After the optimization, $\bsc$ was computed from the true channel (perfect CSI). We swept $L\in\{1,\ldots,10,15,20,25\}$ (row A) and SNR $\in\{-20,-10,\ldots,40\}$\,dB (row B), averaging over six seeds. The targets $\Target$ were the supervised linear equalizer (Linear, ridge regression with $\reg=10^{-3}$ and $4200$ centroid-selected semantic pilots per class) and the zero-shot PPFE (Section~\ref{sec:optimization}) with $\nclusters=20$ prototype clusters ($k$-means) and $\nsamples=1000$ samples per cluster. We report two baselines: \textit{No Mismatch} (upper bound, alignment over a noiseless identity channel) and \textit{Oracle} (the target operator applied digitally to the RX embeddings, without any SIM).

%\noindent 
\textbf{Effect of SIM depth.} Row A of Fig.~\ref{fig:layers_snr} reports the accuracy versus $L$ at infinite SNR. For all schemes, accuracy increases with $L$ until it saturates, confirming that additional degrees of freedom improve the semantic emulation: for instance, the joint scheme with PPFE goes from $10.6\%$ at $L=1$ to $46.2\%$ at $L=3$, $80.0\%$ at $L=10$ and $81.1\%$ at $L=25$, against $93.1\%$ for No Mismatch and $89.6\%$ for Oracle. The joint constructive scheme is the best at small depths (with PPFE, $46.2\%$ against $26.6\%$ for the disjoint symmetric one at $L=3$) and, with PPFE, at every depth, where it reaches the highest accuracy among the dual-SIM schemes ($81.5\%$ at $L=15$), thanks to the end-to-end gradient through $\H$. With Linear, the disjoint symmetric constructive scheme is practically indistinguishable from it beyond $L=3$ (slightly exceeding it, $78.9\%$ against $76.4\%$ at $L=25$), although both end SIMs are trained independently; this is due to its objectives weighting the channel singular modes $\SigHsq$ and following the SVD factorization of $\Target$, so that the two SIMs learn complementary sub-tasks. Among the asymmetric schemes, the equalized variant is the hardest, since one SIM must emulate the full pseudo-inverse $\PseudoInv$; the constructive variant, which includes $\H$ in the per-SIM loss, alleviates this, with its RX version (the TX-SIM compensates the full $384\times384$ channel) being the weakest (below $0.3$ even at $L=25$: $22.1\%$ for Linear, $26.9\%$ for PPFE).

Constructive schemes outperform their equalized counterparts, showing that encoding the channel in the targets is key when the SIM at TX and RX are not co-optimized. At equal $L$, the mono-SIM schemes lag behind the joint and symmetric ones (at $L=10$ with PPFE, $68.7\%$ for the TX mono-SIM against $80.0\%$ for the joint scheme), since a single SIM cannot absorb both the channel and the alignment mapping without deeper stacks, but they outperform the asymmetric ones. At an equal \emph{total} number of layers ($L$ for dual, $2L$ for mono), the gap narrows: for Linear, the joint and symmetric schemes stay ahead (the TX mono-SIM reaches $53.1\%$ at $L=10$, against $62.3\%$ for the joint scheme at $L=5$), whereas, for PPFE, the mono-SIM at $2L$ is comparable to or above the dual schemes at $L$ ($68.7\%$ against $65.9\%$). Moreover, the joint and symmetric schemes use a $20\times20$ channel, about $10\times$ fewer antennas per side than the mono and asymmetric ones ($\Nt=192$ or $384$). PPFE reaches or exceeds Linear at sufficient depth, although Oracle PPFE lies below Oracle Linear: $\AlignMatF$ has rank at most $\nclusters=20$, matching the $20\times20$ channel, and is easier to emulate with phase-only elements than the higher-rank (up to $192$) ridge operator $\AlignMatL$, which the bottleneck truncates.

\begin{figure*}[!t]
  \centering
  \includegraphics[width=\linewidth]{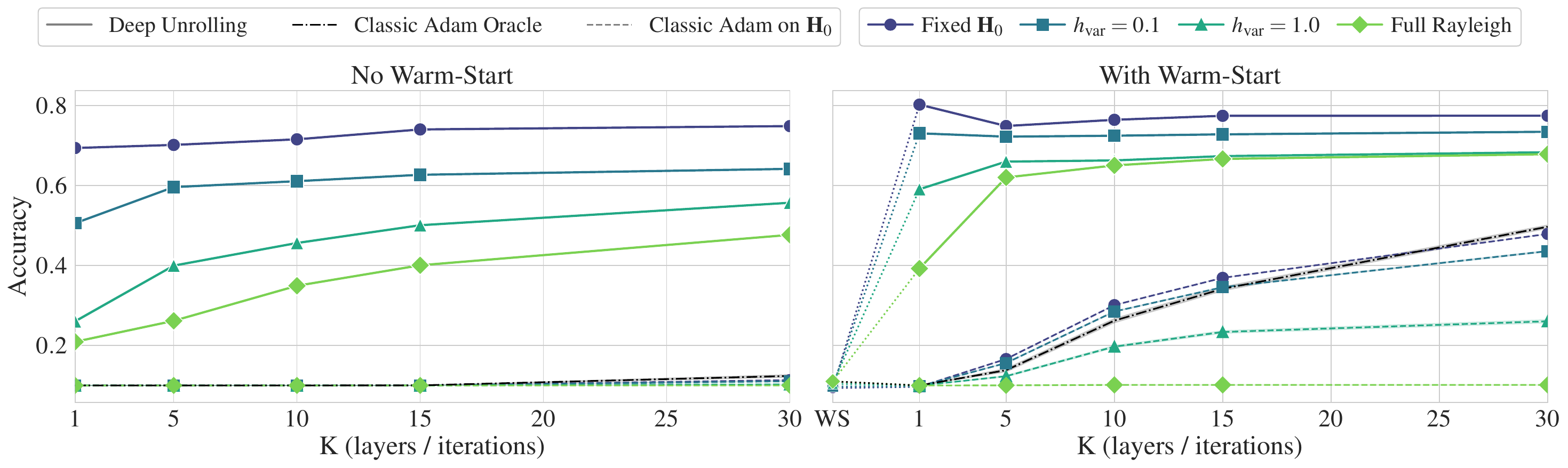}
  \caption{Unrolled network (solid, one marker per depth $K$), $\H_0$-channel baseline (dashed), and oracle (dash-dotted black) for the PPFE target with $L=10$, $20\!\times\!20$ channel and infinite SNR; colors denote the channel distribution. WS: accuracy of the warm-start phases alone. Bands: $95\%$ CI over the test channels.}
  \label{fig:unrolled}
\end{figure*}

%\noindent
\textbf{Effect of SNR.} Row B of Fig.~\ref{fig:layers_snr} reports the accuracy versus SNR$_{[\mathrm{dB}]}$ with $L=10$ SIM layers. All schemes degrade gracefully at low SNR and, from $20\,\text{dB}$ onwards, converge to the saturation accuracy of row A at $L=10$. With PPFE, the joint constructive scheme keeps the highest accuracy over the whole range, with a clear margin over the disjoint symmetric one. With Linear, these two schemes are practically indistinguishable, the symmetric being slightly better at high SNR. The asymmetric schemes remain far below at every SNR, with the equalized variants close to chance level and the constructive ones saturating at low accuracy.

\subsection{Performance of Deep Unrolling}
\label{sec:exp_unrolling}

%\noindent
\textbf{Experiment-specific setup.}
We considered the joint constructive scheme with the same setup as Experiment~A ($L=10$, $20\times20$ channel, PPFE target) and no noise. The classic optimizer solves the problem for one channel and is rerun whenever it changes, whereas the unrolled network is trained once on a known channel distribution and applied to unseen realizations with $K$ unrolled layers. The network followed Algorithm~\ref{alg:unrolled} with diagonal coupling and learnable momentum and scaling ($\stepw^k=0.1$, $\mom^k=0.9$, $\scal^k=\mathbf{1}$). One network was trained for each $K\in\{1,5,10,15,30\}$ with the progressive strategy of Section~\ref{sec:deep_unrolling}, for $100$ epochs, with uniform deep supervision ($\dsratio=1$), Adam with constant learning rates $10^{-3}$, $5\times10^{-4}$, $5\times10^{-4}$, $3\times10^{-4}$, and $10^{-4}$ for the five depths, batch $32$, and gradient clipping at $1$. Training started from the warm start of Section~\ref{sec:deep_unrolling}, computed once ($150$ cycles of $100$ Adam steps per SIM); the network without warm start with all phases initialized at $\pi$ is also reported.

%\noindent 
\textbf{Channel model and evaluation.}
Channels followed a \emph{Rayleigh-plus-mean} (RPM) model: a base channel $\H_0\in\mathbb{C}^{20\times20}$ with i.i.d.\ $\mathcal{CN}(0,1)$ entries was drawn once, and each realization was $\H_i = \H_0 + \sqrt{h_{\mathrm{var}}}\,\chpert_i$ with $[\chpert_i]_{m,n} \sim \mathcal{CN}(0,1)$ i.i.d., where $h_{\mathrm{var}}\ge0$ controls the variability around $\H_0$ ($h_{\mathrm{var}}=0$: fixed channel). We considered $h_{\mathrm{var}}\in\{0,0.1,1\}$ and the pure Rayleigh case, training and testing a separate network for each value. Each dataset had 10\,000 realizations (70\% train / 15\% validation / 15\% test). On each test channel, after each layer (cycle, for the baselines), we formed the cascade $\GR\H\GT$, computed $\bsc$ in closed form from the true channel, aligned the test embeddings, and classified them with the RX classifier of Experiment~A; accuracy was averaged across test channels.

%\noindent 
\textbf{Baselines.} Both baselines ran Adam with one step per SIM and cycle, $\bsc$ recomputed before each step, learning rates $0.1$ (TX) and $0.2$ (RX), from zero-initialized variables or from the same warm start. The \emph{oracle} fitted the optimizer separately on each test channel; since it does not use the channel distribution, it was run on pure Rayleigh channels only and is shown as a single curve. The \emph{$\H_0$-channel} baseline was fit once on $\H_0$ and applied unchanged to the test channels, so it depended on $h_{\mathrm{var}}$ ($\H_0$ was the mean of the distribution, except in the pure Rayleigh case where it was a single realization). Cost was measured in iterations: a cycle of the baselines (one Adam step on the TX and the RX) or a layer of the unrolled network.
%\noindent
\textbf{Results.}
Figure~\ref{fig:unrolled} reports the accuracy averaged over test channels versus the number of iterations or unrolled layers, for different levels of channel variability $h_{\mathrm{var}}$, comparing the deep-unrolled network with the oracle and the $\H_0$-channel baseline, with and without the semantic warm start. Unless otherwise stated, we refer to the warm-started configurations. Deep unrolling attains competitive accuracy with substantially fewer updates than conventional optimization. On pure Rayleigh channels, the oracle reaches $26\%$ and $50\%$ accuracy after $10$ and $30$ iterations, respectively, whereas the unrolled network achieves $62\%$ with only $K=5$ layers and $68\%$ with $K=30$, outperforming the oracle at $30$ iterations by $12$ and $18$ percentage points (pp).

The advantage is more pronounced when the baseline is optimized for a fixed nominal channel $\H_0$ rather than adapted to each realization. After $30$ iterations, this baseline achieves $48\%$ accuracy for $h_{\mathrm{var}}=0$, but decreases to $44\%$ and $26\%$ for $h_{\mathrm{var}}=0.1$ and $1$, and reaches chance level on pure Rayleigh channels. By comparison, the unrolled network trained on each channel distribution achieves $77\%$, $73\%$, $67\%$, and $67\%$, respectively, with $K=15$. The similar results for $h_{\mathrm{var}}=1$ and pure Rayleigh channels reflect the dominance of the random perturbation.

The semantic warm start substantially accelerates adaptation. Although the warm-start phases alone yield only chance-level accuracy ($\approx10\%$) on pure Rayleigh channels, the unrolled network reaches $39\%$ at $K=1$ and $62\%$ at $K=5$ from this initialization, compared with $21\%$ and $26\%$ when all phases start at $\pi$, and only $48\%$ even at $K=30$. Similarly, without warm start the oracle remains near chance level for the first $20$ iterations and reaches only $12\%$ after $30$.

Regarding depth, at low channel variability ($h_{\mathrm{var}}\le0.1$) a single layer already attains near-peak accuracy: for $h_{\mathrm{var}}=0.1$, accuracy remains at $73\%$ for both $K=1$ and $K=15$, while for the fixed channel it decreases slightly from $80\%$ to $77\%$. This is also reflected in the training NMSE, which increases with $K$ in these cases: additional layers do not necessarily help the task and make the optimization harder. Depth is instead beneficial at higher variability, where accuracy increases from $59\%$ at $K=1$ to $67\%$ at $K=15$ for $h_{\mathrm{var}}=1$, and from $62\%$ at $K=5$ to $68\%$ at $K=30$ on pure Rayleigh channels. The warm start and the training are performed once offline, and their cost is amortized over all subsequent channel realizations.
\section{Conclusions}
\label{sec:conclusion}

This paper introduced a dual-SIM framework for OTA semantic alignment between heterogeneous transmitter and receiver models. By jointly exploiting the TX-SIM, the wireless channel, and the RX-SIM, the proposed architecture realizes a target semantic mapping directly in the wave domain, integrating alignment and channel compensation without a dedicated digital equalizer. The framework accommodates supervised linear and zero-shot Parseval-frame targets, and encompasses mono-SIM, disjoint, and joint alignment configurations. The presented numerical results showcased the symmetric disjoint and joint constructive designs achieve the highest accuracy, enabling alignment over compact $20\times20$ MIMO channels with approximately $80\%$ task accuracy at high SNR. The configurations exhibit distinct trade-offs: mono-SIM designs minimize hardware complexity but require larger antenna arrays or deeper stacks, asymmetric disjoint designs yield limited accuracy, symmetric disjoint designs enable decentralized training with a moderate accuracy loss, and the joint design attains the highest accuracy with shallow stacks at the cost of TX--RX coordination; its advantage over a single SIM diminishes at equal total depth. For joint optimization, we developed an alternating projected gradient algorithm with first-order stationarity guarantees and a deep-unrolled network with learnable per-coordinate updates and momentum. The latter enables rapid adaptation to varying channels, outperforming per-channel optimization with substantially fewer updates, especially when initialized through the proposed semantic warm start. Future work will address discrete phase responses, hardware impairments, experimental validation, and jointly learned physical--semantic transceivers.

\appendices
\section{Proof of Theorem~\ref{thm:conv_dualsim}}
\label{app:conv_proof}

% Let $\phv^k=(\phv_{\Tside}^k,\phv_{\Rside}^k)$
% and define
% $F^k\triangleq F(\phv^k,\bsc^k)
% =\bar F(\phv^k)$,
% since $\bsc^k$ minimizes $F$ over $\bsc$
% for the current phases.
% Under the assumption $z_{\min}>0$,
% the optimal scaling $\bsc(\phv)$ is
% continuous and bounded on $\phset$. Thus, the block gradients of $F$ satisfy
% the Lipschitz conditions stated in Theorem~\ref{thm:conv_dualsim}. 

In the sequel, we prove Points 1 and 2 of Theorem~\ref{thm:conv_dualsim}.

\emph{Point 1).} Define the phase increments $\boldsymbol{\Delta}_s^k \triangleq \phv_s^{k+1}-\phv_s^k$, for $s\in\{\Tside,\Rside\}$. By the optimality condition of Euclidean projection onto a closed convex set~\cite[Prop.~2.1.3]{Bertsekas16}, the TX update satisfies the condition:
\begin{equation}
\big\langle \nabla_{\phv_{\Tside}}
F(\phv_{\Tside}^k,\phv_{\Rside}^k,\bsc^k),
\boldsymbol{\Delta}_{\Tside}^k \big\rangle
\le
-\frac{1}{\lr_{\Tside}}
\|\boldsymbol{\Delta}_{\Tside}^k\|_2^2.
\label{eq:proj_descent_tx}
\end{equation}
Combining this inequality with the descent lemma for Lipschitz-continuous gradients (see, e.g.,~\cite{Beck17}) yields:
\begin{equation}
F(\phv_{\Tside}^{k+1},\phv_{\Rside}^k,\bsc^k)
\le F^k-c_{\Tside}
\|\boldsymbol{\Delta}_{\Tside}^k\|_2^2,
\label{eq:descent_lemma_tx}
\end{equation}
where
$c_{\Tside}\triangleq
1/\lr_{\Tside}-\lip_{\Tside}/2>0$. The same argument applies to the RX update, evaluated at the updated TX phases while keeping $\bsc^k$ fixed. Thus, the following condition holds:
\begin{equation}
F(\phv_{\Tside}^{k+1},\phv_{\Rside}^{k+1},\bsc^k)
\le
F(\phv_{\Tside}^{k+1},\phv_{\Rside}^k,\bsc^k)
-c_{\Rside}\|\boldsymbol{\Delta}_{\Rside}^k\|_2^2
\label{eq:descent_lemma_rx}
\end{equation}
with $c_{\Rside}\triangleq 1/\lr_{\Rside}-\lip_{\Rside}/2>0$.  Finally, since $\bsc^{k+1}$ minimizes $F(\phv^{k+1},\bsc)$, we have $F^{k+1}\le F(\phv^{k+1},\bsc^k)$. Combining~\eqref{eq:descent_lemma_tx} and~\eqref{eq:descent_lemma_rx} with this exact scaling update, yields:
\begin{equation}
F^{k+1}
\le F^k
-c_{\Tside}\|\boldsymbol{\Delta}_{\Tside}^k\|_2^2
-c_{\Rside}\|\boldsymbol{\Delta}_{\Rside}^k\|_2^2.
\label{eq:three_block_descent}
\end{equation}
Since $F^k=\bar F(\phv^k)\ge0$, the objective sequence is non-increasing and convergent, proving the first claim.

\emph{Point 2).} Summing~\eqref{eq:three_block_descent} over $k$ and using $F^k\ge0$ and $c_{\Tside},c_{\Rside}>0$, yields:
\begin{equation}
\sum_{k=0}^{\infty}
\left(
\|\boldsymbol{\Delta}_{\Tside}^k\|_2^2+
\|\boldsymbol{\Delta}_{\Rside}^k\|_2^2
\right)<\infty.
\label{eq:summability}
\end{equation}
Hence, $\boldsymbol{\Delta}_{\Tside}^k\to0$ and $\boldsymbol{\Delta}_{\Rside}^k\to0$. 
% Let $\phv^\star$ be an arbitrary limit point
% of $\{\phv^k\}$, and consider a subsequence
% $\phv^{k_j}\to\phv^\star$, which exists by
% compactness of $\phset$.
% By~\eqref{eq:summability}, we also have
% $\phv^{k_j+1}\to\phv^\star$.
% Moreover, the continuity of the optimal
% scaling under $z_{\min}>0$ ensures
% $\bsc^{k_j}\to
% \bsc^\star\triangleq\bsc(\phv^\star)$. 
% By compactness of $\phset$, the sequence
% $\{\phv^k\}$ admits a convergent subsequence
% $\phv^{k_j}\to\phv^\star$.
% Equation~\eqref{eq:summability} implies that
% $\phv^{k_j+1}\to\phv^\star$ as well.
% Since $z_{\min}>0$, the optimal scaling
% is continuous, yielding
% $\bsc^{k_j}\to\bsc^\star
% \triangleq\bsc(\phv^\star)$.
% Since the gradients of $F$ and the Euclidean
% projections are continuous, taking the limit
% in the alternating updates
% \eqref{eq:xiT_update} and \eqref{eq:xiR_update}
% gives:
% \begin{equation}
% \phv_s^\star=
% \Pi_{\phset_s}\left(
% \phv_s^\star-
% \lr_s\nabla_{\phv_s}
% F(\phv^\star,\bsc^\star)
% \right),
% \quad s\in\{\Tside,\Rside\}.
% \label{eq:proj_stationarity}
% \end{equation}
% Although the RX gradient uses the scaling
% computed before the TX update, this does not
% affect~\eqref{eq:proj_stationarity}, since
% both phase iterates converge to the same
% limit along the subsequence. 
Since $\phset$ is compact, the sequence $\{\phv^k\}$ admits at least one limit point. Consider any such point $\phv^\star$ and a subsequence $\phv^{k_j}\to\phv^\star$. By~\eqref{eq:summability}, the difference between consecutive iterates vanishes, so $\phv^{k_j+1}\to\phv^\star$ as well. Furthermore, since $z_{\min}>0$, the optimal scaling $\bsc(\phv)$ is continuous, which implies $\bsc^{k_j}\to\bsc^\star \triangleq\bsc(\phv^\star)$. We can therefore pass to the limit in the projected-gradient updates \eqref{eq:xiT_update} and \eqref{eq:xiR_update}. By continuity of the gradients and Euclidean projections, both updates yield the fixed-point conditions
\begin{equation}
\phv_s^\star=
\Pi_{\phset_s}\left(
\phv_s^\star-
\lr_s\nabla_{\phv_s}
F(\phv^\star,\bsc^\star)
\right),
\quad s\in\{\Tside,\Rside\}.
\label{eq:proj_stationarity}
\end{equation}
In particular, the RX update uses the new TX phases but the scaling computed before the TX update. This does not affect the limit, since both the old and updated TX phases converge to $\phv_\Tside^\star$, while the scaling converges to $\bsc^\star$.  By the optimality condition for Euclidean projection onto a closed convex set (see~\cite[Prop.~2.1.3]{Bertsekas16}), \eqref{eq:proj_stationarity} implies
\begin{equation}
\nabla_{\phv_s}F(\phv^\star,\bsc^\star)^\top
(\phv_s-\phv_s^\star)\ge0,
\quad \forall\phv_s\in\phset_s.
\label{eq:stationarity_vi}
\end{equation}
Since $\bsc^\star$ minimizes $F(\phv^\star,\cdot)$, the derivative through the optimal scaling vanishes, yielding $\nabla_{\phv_s}\bar F(\phv^\star) =\nabla_{\phv_s}F(\phv^\star,\bsc^\star)$. Thus, every limit point $\phv^\star$ is first-order stationary for $\bar F$ over $\phset$, with $\bsc^\star$ being the corresponding optimal scaling.
\hfill$\square$

\bibliographystyle{IEEEtran}
\bibliography{refs}

\newpage
\makeatletter
\newcommand{\suppfigcaption}[1]{\def\@captype{figure}\caption{#1}}
\makeatother
\setcounter{dbltopnumber}{2}

% ------------------------------ PAGE 1 --------------------------------
\twocolumn[{
\centering
\includegraphics[width=0.96\textwidth]
{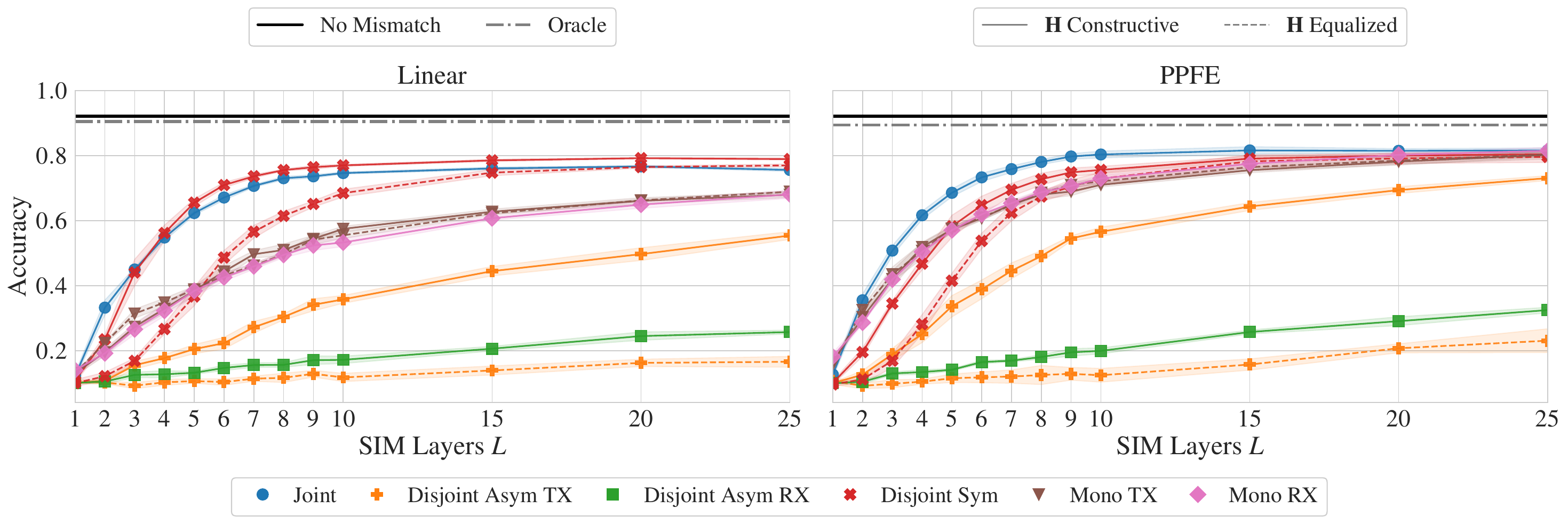}
\suppfigcaption{Linear-probe accuracy versus the number of layers per SIM $L$ for the Linear (left) and PPFE (right) targets, considering $32\times32$ meta-atoms per layer and infinite SNR. Channel dimensions: $20\times20$ (joint and disjoint symmetric), $192\times192$ (mono TX and disjoint asymmetric TX), and $384\times384$ (mono RX and disjoint asymmetric RX). Shaded regions: $95\%$ CI over six seeds.}
\label{fig:supp_layers}
\vspace{6pt}
\includegraphics[width=0.96\textwidth]
{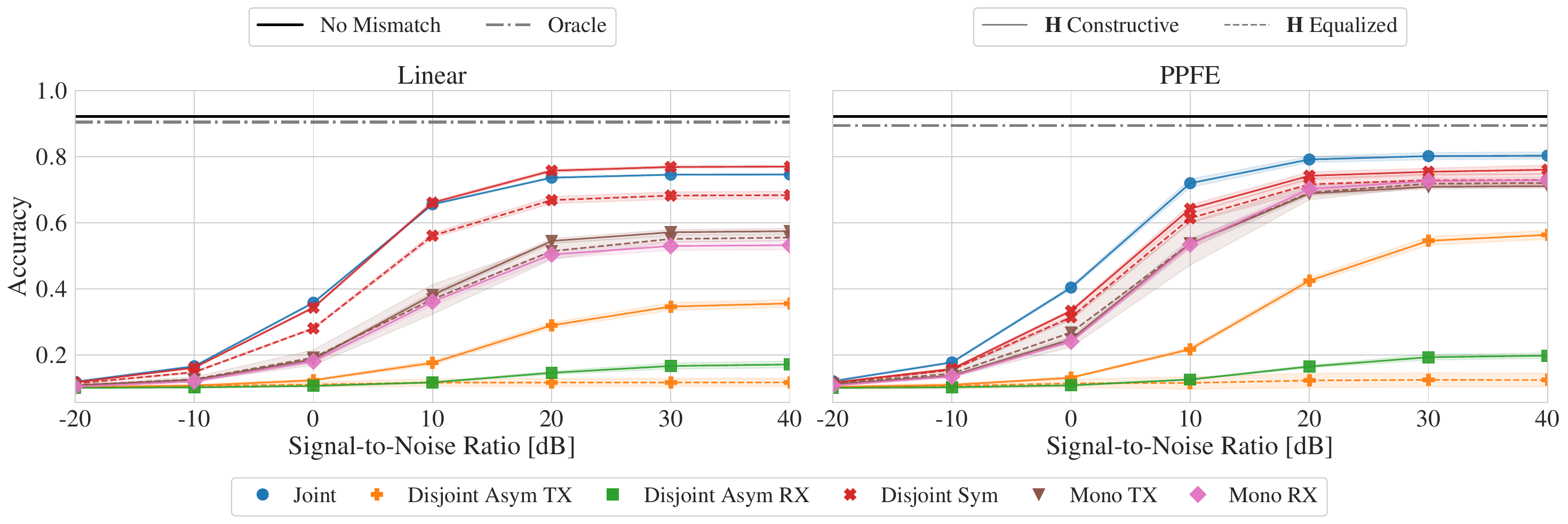}
\suppfigcaption{Linear-probe accuracy versus SNR for the Linear (left) and PPFE (right) targets with $L=10$ and channel dimensions as in Fig.~\ref{fig:supp_layers}. Shaded regions: $95\%$ CI over six seeds.}
\label{fig:supp_snr}
\vspace{4pt}
}]

\section*{Supplementary Material}
\label{sec:supplementary}

% ------------------------------ PAGE 2 --------------------------------
\begin{figure*}[t]
\centering
\includegraphics[width=0.96\textwidth]%
{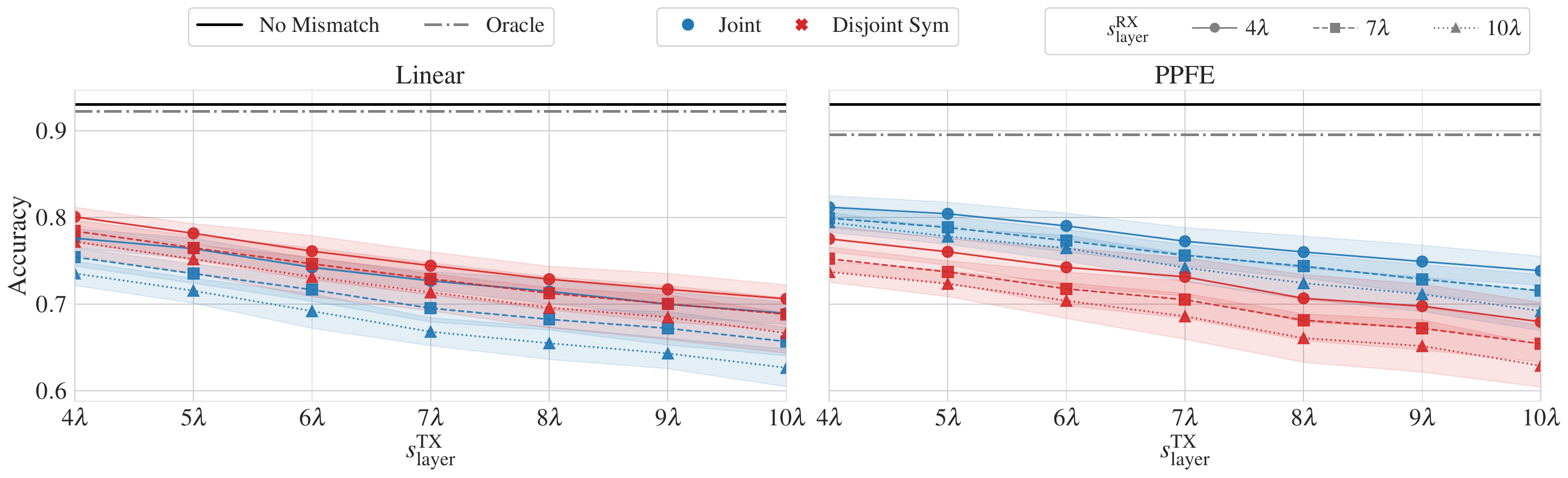}
\caption{MLP classification accuracy versus TX-SIM spacing $s_{\mathrm{layer}}^{\Tside}$ for Linear (left) and PPFE (right) alignment, considering RX spacing $s_{\mathrm{layer}}^{\Rside} \in\{4,7,10\}\wavelength$ (line style), $L=10$, a $20\times20$ channel, and infinite SNR. Shaded regions: $95\%$ CI over six seeds.}
\label{fig:supp_slayer_neural}
\end{figure*}

\begin{figure*}[t]
\centering
\includegraphics[width=0.96\textwidth]%
{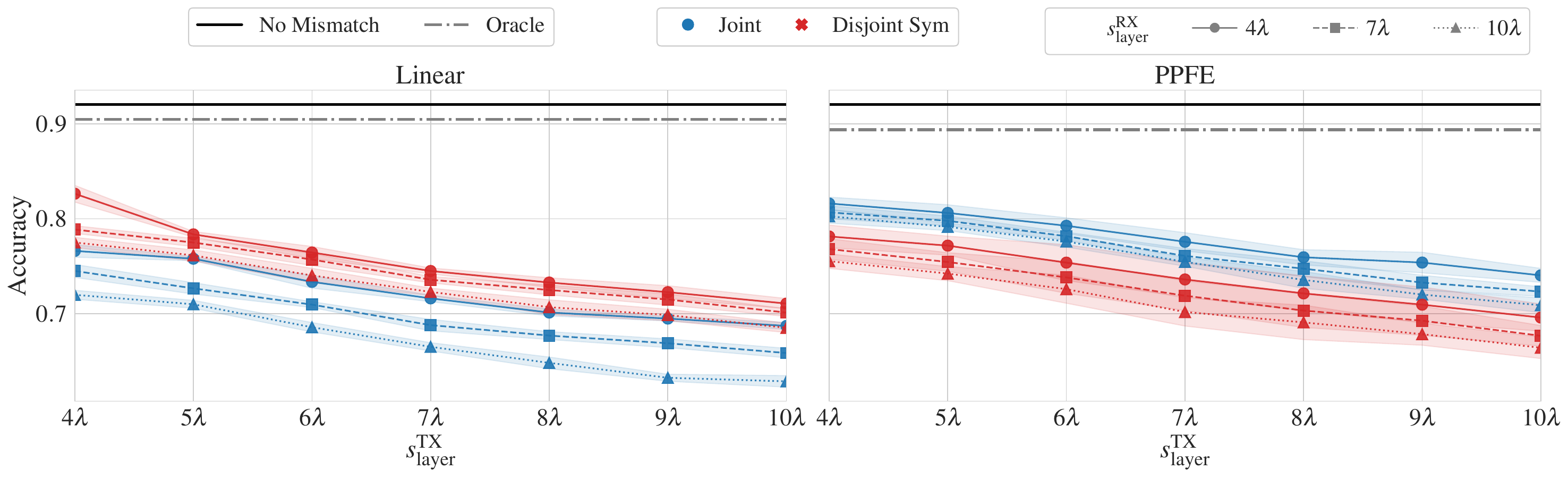}
\caption{Linear-probe accuracy versus TX-SIM spacing under the same conditions as Fig.~\ref{fig:supp_slayer_neural}. Shaded regions: $95\%$ CI over six seeds.}
\label{fig:supp_slayer_probe}
\end{figure*}

% ------------------------------ PAGE 3 --------------------------------
\begin{figure*}[t]
\centering
\includegraphics[width=0.96\textwidth]%
{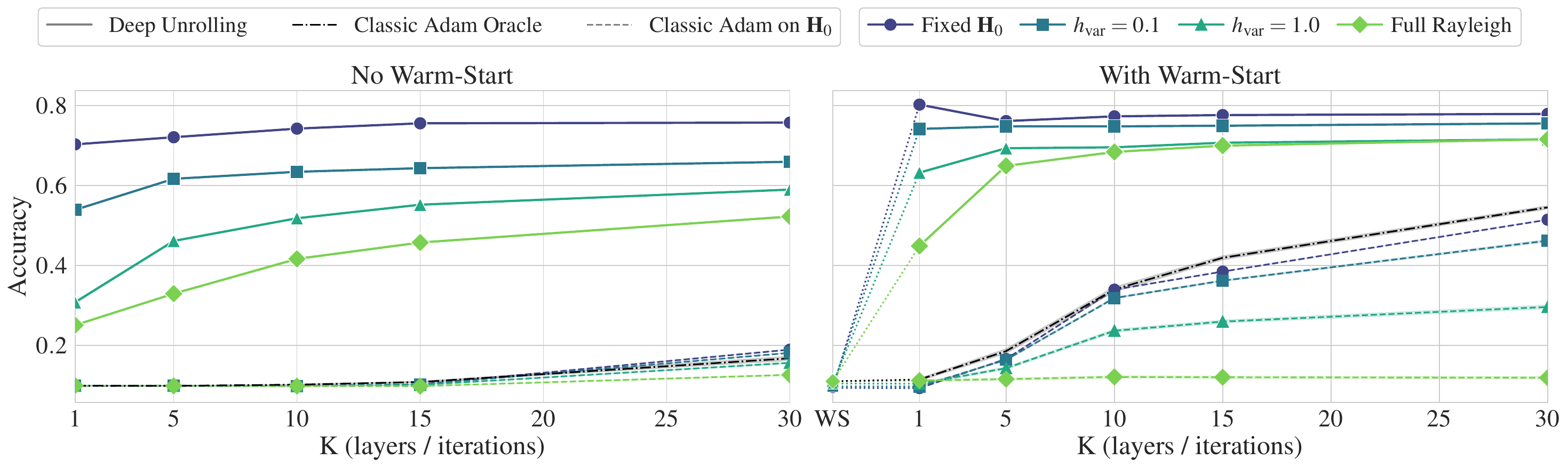}
\caption{Linear-probe accuracy of the proposed deep-unrolled network (solid), the $\H_0$-optimized baseline (dashed), and the per-channel oracle (black dash-dotted) for PPFE, $L=10$, a $20\times20$ channel, and infinite SNR. Left: without semantic warm start; right: with warm start. Colors indicate channel distributions and WS denotes the accuracy of the warm-start configuration alone. Shaded regions: $95\%$ CI over test channels.}
\label{fig:supp_unrolled}
\end{figure*}

\begin{table*}[t]
\centering
\caption{Accuracy (\%) of the proposed unrolled network ($K=30$) and conventional baselines run for $10^4$ iterations.\\ Peak denotes the highest observed accuracy and its iteration (in parentheses). The $\H_0$ baseline is fitted on the nominal channel and applied to test channels; the oracle is fitted separately on each test channel (shown for pure Rayleigh fading channels).}
\label{tab:supp_conv}
\footnotesize
\setlength{\tabcolsep}{4pt}
\begin{tabular}{llcccccc}
\toprule
& & \multicolumn{3}{c}{Linear probe}
& \multicolumn{3}{c}{MLP}\\
\cmidrule(lr){3-5}
\cmidrule(lr){6-8}
Baseline & Channel & Unrolled
& Peak (iteration) & At $10^4$
& Unrolled & Peak (iteration) & At $10^4$\\
\midrule
\multicolumn{8}{l}{\emph{With semantic warm start}}\\
$\H_0$ & $h_{\mathrm{var}}=0$
& 77.9 & 80.3 (8605) & 79.1
& 77.4 & 80.0 (9431) & 78.3\\
$\H_0$ & $h_{\mathrm{var}}=0.1$
& 75.6 & 60.4 (107) & 46.8
& 73.4 & 56.6 (212) & 45.0\\
$\H_0$ & $h_{\mathrm{var}}=1$
& 71.6 & 32.5 (81) & 20.1
& 68.3 & 26.7 (42) & 15.7\\
$\H_0$ & Rayleigh
& 71.6 & 12.1 (9) & 11.4
& 67.8 & 10.1 (18) & 10.0\\
Oracle & Rayleigh
& 71.6 & 79.5 (9988) & 79.4
& 67.8 & 79.0 (9908) & 79.0\\
\midrule
\multicolumn{8}{l}{\emph{Without semantic warm start}}\\
$\H_0$ & $h_{\mathrm{var}}=0$
& 75.8 & 80.2 (9964) & 79.2
& 74.9 & 80.0 (9469) & 78.6\\
$\H_0$ & $h_{\mathrm{var}}=0.1$
& 66.0 & 57.6 (597) & 43.5
& 64.2 & 55.3 (469) & 40.5\\
$\H_0$ & $h_{\mathrm{var}}=1$
& 59.0 & 28.7 (439) & 19.4
& 55.7 & 24.4 (187) & 13.8\\
$\H_0$ & Rayleigh
& 52.3 & 14.2 (40) & 11.4
& 47.7 & 10.7 (54) & 10.0\\
Oracle & Rayleigh
& 52.3 & 78.8 (9945) & 78.8
& 47.7 & 78.5 (9966) & 78.5\\
\bottomrule
\end{tabular}
\end{table*}

% ------------------------------ PAGE 4 --------------------------------
\begin{figure*}[t]
\centering
\begin{minipage}[t]{0.49\textwidth}
\centering
\includegraphics[width=\linewidth]%
{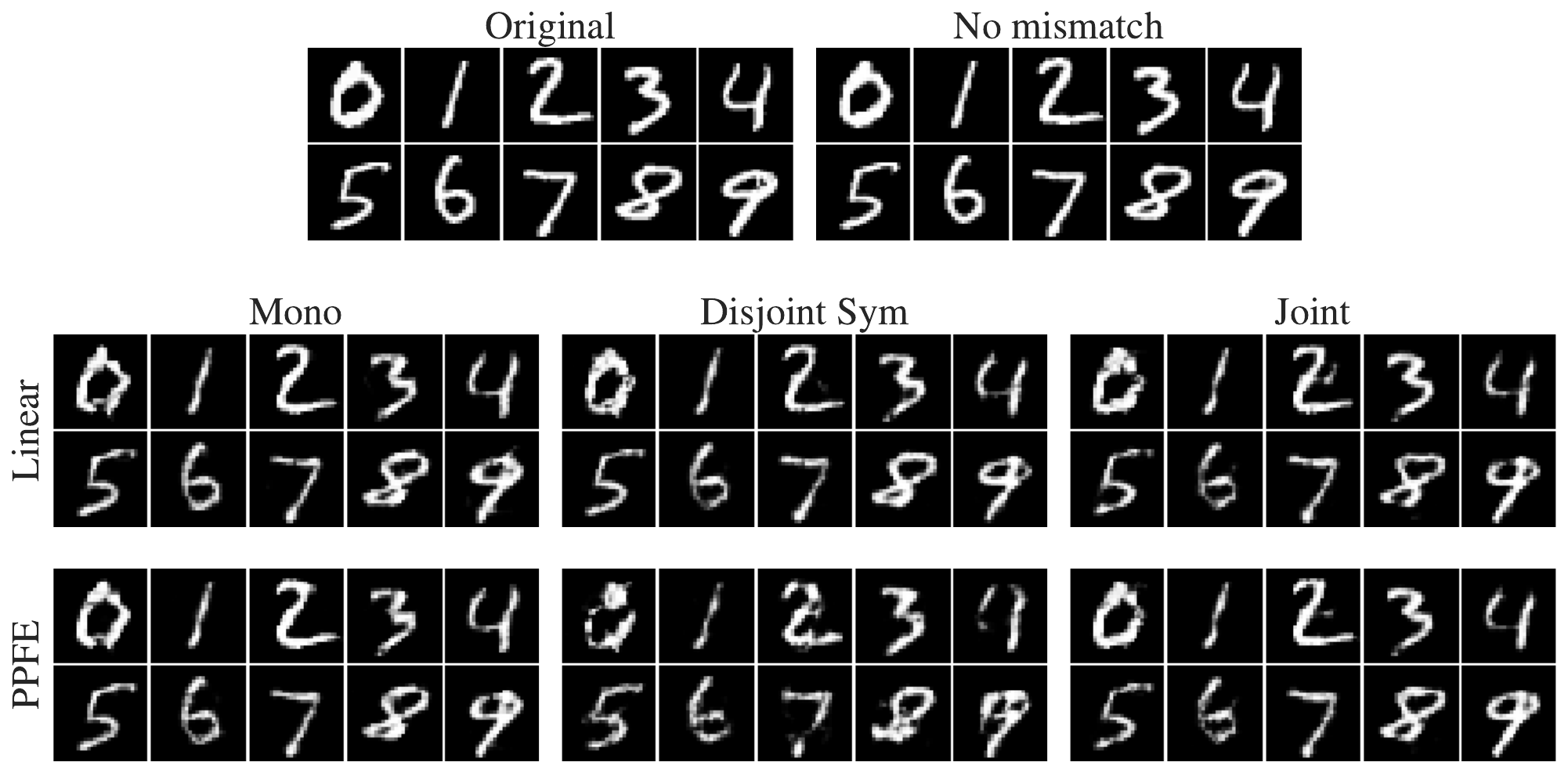}
\par\small (a) MNIST
\end{minipage}\hfill
\begin{minipage}[t]{0.49\textwidth}
\centering
\includegraphics[width=\linewidth]%
{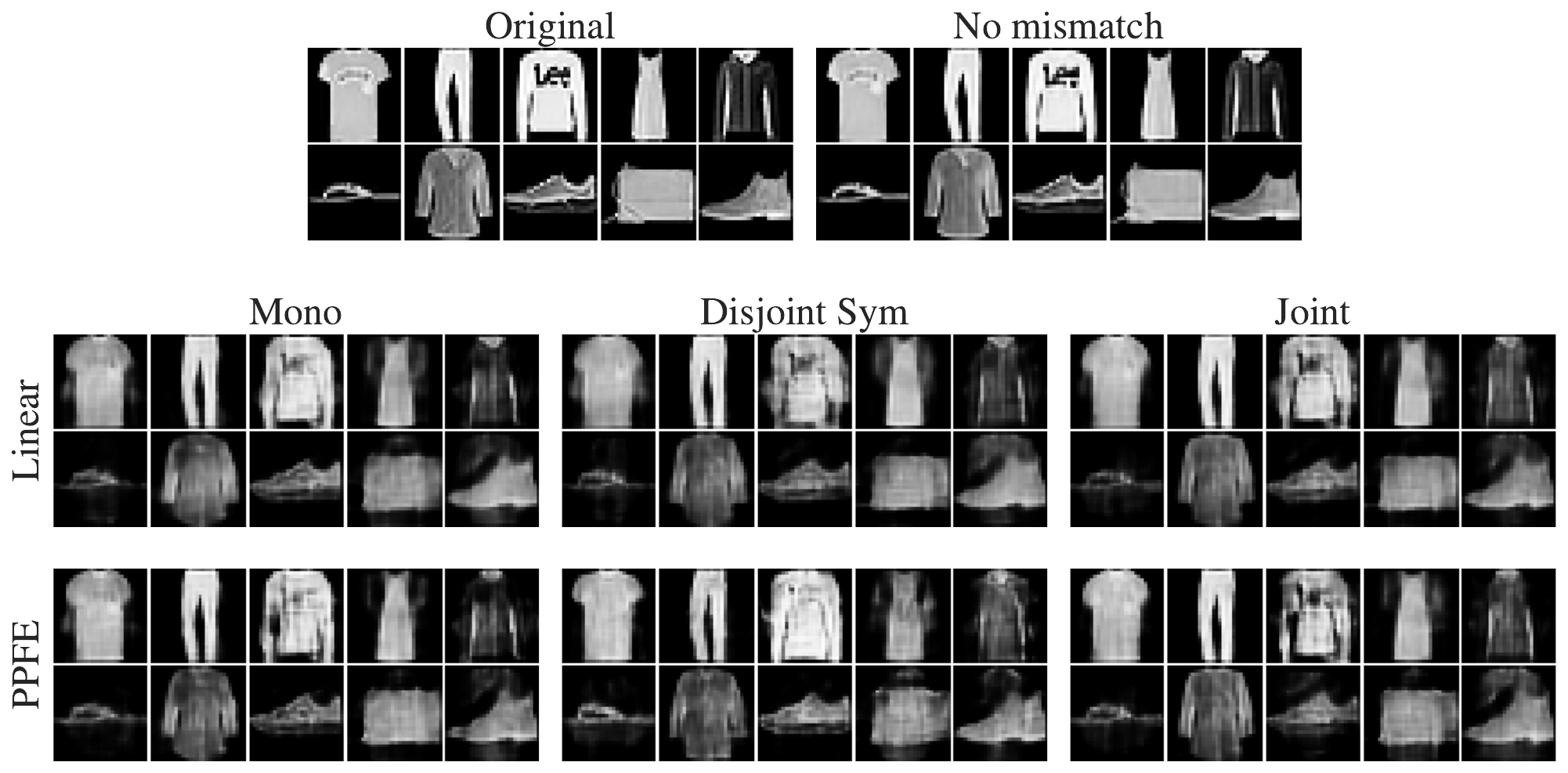}
\par\small (b) Fashion-MNIST
\end{minipage}
\caption{Example test-image reconstructions for MNIST and Fashion-MNIST (one sample per class). The top rows show the original images and the RX decoder outputs from its own embeddings (No Mismatch). The remaining rows compare mono-SIM, disjoint symmetric, and joint constructive alignment for the Linear and PPFE targets. All schemes use $L=50$, $64\times64$ meta-atoms per layer, and infinite SNR.}
\label{fig:supp_rec}
\end{figure*}

This section complements the numerical evaluation in Section~\ref{sec:numerical} by providing additional results on linear-probe classification, the impact of SIM inter-layer spacing, and the convergence of conventional optimization baselines. We also extend the evaluation to image reconstruction, assessing the applicability of the proposed OTA semantic alignment framework beyond classification.
 The numerical study in Section~\ref{sec:numerical} assessed semantic alignment through a two-layer MLP trained on the original RX embeddings and then kept fixed. To examine whether the reported trends depend on the evaluation classifier, we also used a linear probe, fitted by ordinary least squares without a bias term. Let $\mathbf{S}\in\mathbb{R}^{N\times\Rdim}$ contain the RX embeddings of the training split and $\mathbf{Y}\in\{0,1\}^{N\times10}$ their one-hot labels. The probe weights solve the problem: 
\begin{equation}
\mathbf{W}^{\star}
\triangleq\underset{\mathbf{W}\in
\mathbb{R}^{\Rdim\times10}}{\arg\min}
\;\|\mathbf{S}\mathbf{W}-\mathbf{Y}\|_{\mathrm F}^{2}.
\label{eq:supp_probe_fit}
\end{equation}
For an aligned RX embedding
$\hat{\mathbf{s}}_{\Rside}$, the predicted
class is
$\arg\max_c[\hat{\mathbf{s}}_{\Rside}^{\top}
\mathbf{W}^{\star}]_c$.
For each encoder pair, the probe was fitted once using the same original RX representations employed to train the MLP. It was subsequently held fixed and applied to the aligned embeddings, including those of the \emph{No Mismatch} and \emph{Oracle} baselines. Unlike the MLP, least-squares fitting has no stochastic training variability, although evaluation results can still vary across channel and SIM realizations.

\subsection{Additional Results for Experiment A}
\label{sec:supp_expA}

Figures~\ref{fig:supp_layers} and~\ref{fig:supp_snr} report linear-probe classification accuracy versus the number of layers per SIM and the SNR, respectively. They complement the corresponding MLP-based comparisons of Section~\ref{sec:exp_arch} with the same experimental setup and the probe in \eqref{eq:supp_probe_fit}. The \emph{No Mismatch} baseline attains $92.1\%$ accuracy, while the \emph{Oracle} achieves $90.4\%$ with the Linear target and $89.4\%$ with PPFE. The relative performance of the SIM configurations is consistent with the MLP evaluation.

%\noindent
\textbf{Impact of depth.}
Figure~\ref{fig:supp_layers} shows accuracy as a function of $L$ at infinite SNR. As shown, joint constructive optimization is especially effective for shallow stacks. With PPFE, it outperforms the disjoint symmetric constructive scheme throughout the tested range, achieving $50.8\%$ versus $34.5\%$ at $L=3$ and $80.3\%$ versus $75.6\%$ at $L=10$. For the Linear target, the joint scheme leads up to $L=3$, but the disjoint symmetric design overtakes it at $L\ge5$ and reaches $78.9\%$ versus $75.6\%$ at $L=25$. At this depth, the asymmetric constructive TX and RX configurations and the asymmetric equalized TX configuration attain, respectively, $55.4\%$, $25.8\%$, and $16.6\%$ with Linear, compared with $73.1\%$, $32.5\%$, and $23.1\%$ with PPFE. Mono-SIM accuracy at $L=25$ is $68.0$--$68.9\%$ with Linear and $80.2$--$81.5\%$ with PPFE, approaching the dual-SIM designs only in the latter case.

%\noindent
\textbf{Impact of SNR.}
Figure~\ref{fig:supp_snr} shows accuracy versus SNR for $L=10$. It can be observed that accuracy generally improves with SNR and approaches saturation beyond $20$\,dB. For the joint design, Linear accuracy increases from $11.9\%$ at $-20$\,dB to $35.9\%$ at $0$\,dB and $65.6\%$ at $10$\,dB, reaching $73.6$--$74.6\%$ above $20$\,dB. The corresponding PPFE accuracies are $12.1\%$, $40.4\%$, $72.0\%$, and $79.1$--$80.3\%$. At $40$\,dB, the disjoint symmetric design exceeds the joint one by $2.4$ percentage points for Linear, whereas the joint design leads by $4.3$ percentage points for PPFE. Equalized asymmetric configurations remain near chance level ($10\%$), and the constructive asymmetric TX and RX designs saturate at $35.6\%$ and $17.2\%$ with Linear, versus $56.4\%$ and $19.9\%$ with PPFE.

%\noindent
\textbf{Impact of inter-layer spacing.}
\label{sec:supp_spacing} The main experiments used an inter-layer spacing of $s_{\mathrm{layer}}=5\wavelength$. We additionally examined the joint and disjoint symmetric constructive configurations with independently varied TX- and RX-SIM spacings. Both schemes used a $20\times20$ MIMO channel, $L=10$ layers per SIM, infinite SNR, and the optimization settings of Experiment~A. We varied $s_{\mathrm{layer}}^{\Tside} \in\{4,\ldots,10\}\wavelength$ and $s_{\mathrm{layer}}^{\Rside} \in\{4,7,10\}\wavelength$, averaging over six seeds.

\subsubsection{MLP results}
Figure~\ref{fig:supp_slayer_neural} plots MLP accuracy against TX-SIM inter-layer spacing for three RX-SIM spacings and both targets. It shows a consistent decrease in accuracy as the inter-layer spacing increases. Raising the TX spacing from $4\wavelength$ to $10\wavelength$ reduces accuracy by $7.3$--$10.9$ percentage points; for example, joint PPFE alignment falls from $81.2\%$ to $73.9\%$ when $s_{\mathrm{layer}}^{\Rside}=4\wavelength$. Increasing the RX spacing over the same range causes a smaller reduction of $1.7$--$6.3$ percentage points. The greater sensitivity to the TX spacing is consistent with the broader spatial spreading induced by propagation between phase-only layers, which can reduce the independent field patterns available to the cascade. These results favor relatively compact layer spacing, particularly at the TX, within the propagation model considered. The ranking of the two configurations is preserved across the spacing grid. With PPFE, the joint design leads the disjoint symmetric one by $3.7$--$6.3$ percentage points; with Linear, the disjoint symmetric design leads by $1.4$--$4.5$ percentage points.

\subsubsection{Linear-probe results}
Figure~\ref{fig:supp_slayer_probe} reports the corresponding spacing experiment with the linear probe and confirms the trends in Fig.~\ref{fig:supp_slayer_neural}. Increasing TX spacing from $4\wavelength$ to $10\wavelength$ reduces probe accuracy by $7.6$--$11.6$ percentage points, compared with $1.3$--$6.2$ percentage points for the RX spacing. Joint PPFE alignment leads by $3.4$--$5.3$ percentage points throughout the grid, whereas disjoint symmetric Linear alignment leads by $2.4$--$6.6$ percentage points. The corresponding confidence bands extend up to $1.9$ percentage points.

\subsection{Additional Results for Experiment B}
\label{sec:supp_expB}

%\noindent
\textbf{Deep unrolling with a linear probe.}
Figure~\ref{fig:supp_unrolled} shows linear-probe accuracy versus the number of unrolled layers or optimization iterations for different channel distributions, with and without the semantic warm start. It complements the MLP-based evaluation of Section~\ref{sec:exp_unrolling} using the same networks and optimization baselines. The results confirm the advantage of learned, channel-dependent SIM adaptation with few unrolled layers. On pure Rayleigh channels, the per-channel oracle initialized with the semantic warm start attains $34.0\%$, $54.5\%$, and $66.1\%$ after $10$, $30$, and $60$ iterations ($26.2\%$, $49.7\%$, and $63.5\%$ with the MLP). The unrolled network nearly matches the $60$-iteration oracle with $K=5$ ($64.9\%$ with the probe, $62.0\%$ with the MLP, against $66.1\%$ and $63.5\%$) and exceeds it with $K=30$ ($71.6\%$ and $67.8\%$). For the nominal-channel baseline fitted on $\H_0$, $60$ iterations with warm start yield $64.1\%$, $56.1\%$, $32.3\%$, and $11.7\%$ for $h_{\mathrm{var}}=0$, $0.1$, $1$ and pure Rayleigh channels, respectively ($62.0\%$, $53.8\%$, $26.6\%$, and $10.0\%$ with the MLP). By comparison, the distribution-trained unrolled networks achieve $77.6\%$, $75.0\%$, $70.7\%$, and $70.0\%$ with $K=15$ ($77.4\%$, $72.8\%$, $67.4\%$, and $66.7\%$ with the MLP). Thus, the learned updates remain effective under channel variations for which a fixed-channel configuration degrades sharply. The semantic warm start substantially improves early-layer performance. Although the warm-start phases alone yield only $9.3$--$11.0\%$ accuracy across the tested channel distributions, the unrolled network achieves $44.9\%$ at $K=1$ and $64.9\%$ at $K=5$ on pure Rayleigh channels, versus $25.1\%$ and $32.9\%$ without warm start. Even at $K=30$, the latter reaches only $52.3\%$. Without warm start, the oracle remains near $10$--$12\%$ for its first 20 iterations and reaches $32.3\%$ at $60$ iterations. Depth gains are largest at higher variability: for $h_{\mathrm{var}}=1$, accuracy increases from $63.2\%$ at $K=1$ to $70.7\%$ at $K=15$, but remains nearly unchanged for $h_{\mathrm{var}}=0.1$ ($74.2\%$ to $75.0\%$). For a fixed channel, $K=1$ reaches $80.3\%$, compared with $77.6\%$ at $K=15$. These results were obtained with a single training seed, as in the main experiment.

%\noindent
\textbf{Long-run optimization baselines.}
To distinguish optimization speed from the eventual accuracy of conventional iterative methods, we extended their runs to $10^4$ iterations. Table~\ref{tab:supp_conv} reports the best observed classification accuracy (and the iteration at which it occurs), the accuracy after $10^4$ iterations, and the performance of the $K=30$ unrolled model, evaluated with both the linear probe and the MLP. The baseline optimized for $\H_0$ and transferred unchanged to test channels does not match the unrolled model when $h_{\mathrm{var}}\in\{0.1,1\}$ or on pure Rayleigh channels, even after extended optimization. Its test accuracy can decline with the iteration count because further fitting to $\H_0$ does not improve performance on other channels. When $h_{\mathrm{var}}=0$, by contrast, the test channel equals $\H_0$, and long-run accuracy becomes comparable to that of the unrolled model. With warm start, the $\H_0$ baseline first surpasses the $K=30$ unrolled accuracy after 1038 iterations with the probe and 1332 with the MLP. Its maximum values ($80.3\%$ and $80.0\%$, respectively) are comparable to the $K=1$ unrolled results ($80.3\%$ and $80.2\%$). Without warm start, the corresponding crossover requires $2135$ and $3238$ iterations. The per-channel oracle avoids nominal-channel mismatch and eventually exceeds the unrolled model on pure Rayleigh channels. With warm start, it takes $121$ iterations with the probe and $93$ with the MLP to match the $K=30$ unrolled model; without warm start, the counts increase to $668$ and $527$. These are iteration-count comparisons rather than equal-runtime comparisons, but they highlight the tradeoff: the oracle must be reoptimized for each channel, whereas the trained network produces a channel-specific configuration in $K$ unrolled layers.

\subsection{OTA Semantic Alignment for Image Reconstruction}
\label{sec:supp_rec}

The preceding results assess semantic alignment through classification accuracy. We further evaluate whether the same wave-domain framework preserves the information required for \emph{image reconstruction}, where the RX decoder maps the aligned TX latent representation to an image rather than a class label. We used MNIST~[S1] and Fashion-MNIST~[S2], each comprising ten classes of $28\times28$ grayscale images. For each dataset, two independently trained convolutional autoencoders with different widths served as the TX and RX models. Their encoders used three stride-$2$ convolutional blocks (a $3\times3$ convolution, batch normalization, and sigmoid linear unit (SiLU) activation), each followed by a residual block, and a linear projection to the latent space. The decoders employed residual blocks and PixelShuffle upsampling, followed by bilinear resizing to $28\times28$, convolution, and sigmoid output. The TX model used channel widths $[128,256,512]$ and $512$ real latent features ($\Tdimc=256$ complex symbols); the RX model used $[64,128,256]$ and $256$ real features ($\Rdimc=128$ complex symbols). We used the posterior means as latent embeddings. Both autoencoders were trained as MMD-InfoVAEs. The reconstruction objective combined mean squared and $L_1$ errors, a structural-similarity term, and a perceptual term based on early visual geometry group (VGG) features. An additional maximum mean discrepancy penalty matched the aggregate latent distribution to an isotropic Gaussian, without a Kullback–Leibler term, promoting well-conditioned pilot covariance for whitening. Training used Adam with initial learning rates $5\times10^{-4}$ (TX) and $10^{-3}$ (RX), cosine annealing, batches of $256$, and early stopping within $200$ epochs. The supervised Linear operator was obtained by ridge regression with $\gamma=10^{-3}$. For PPFE, anchor prototypes were obtained by $k$-means clustering of the pilots, averaging up to $500$ samples per prototype. Semantic pilots were selected by proximity to class centroids, with at most $5000$ samples per class. The mono-SIM and joint configurations used $2000$ PPFE prototypes; the disjoint symmetric configuration used at most $128$, as its prototype count was constrained by the channel dimension.

We evaluated constructive mono-SIM (TX), disjoint symmetric, and joint constructive designs, using $L=50$ layers per SIM and $64\times64$ meta-atoms per layer. The wavelength, unit-cell dimensions, and inter-layer spacing followed Experiment~A. All configurations used a $128\times128$ flat Rayleigh channel, infinite SNR, and perfect channel state information. Phase logits were initialized to zero and optimized using Adam with cosine learning-rate scheduling. The joint scheme used $150$ outer cycles of $100$ steps per SIM at learning rate $0.03$; mono-SIM and disjoint symmetric schemes used $3000$ steps per SIM at learning rate $0.1$. Results are reported for one random seed. In particular, Fig.~\ref{fig:supp_rec} compares the original images, the no-mismatch RX reconstructions, and the reconstructions obtained with each SIM configuration and semantic target for MNIST and Fashion-MNIST. It presents one test image per class. The results illustrate that features generated by the heterogeneous TX encoder can be mapped into a representation usable by the RX image decoder. For MNIST, all configurations preserve recognizable digits and their main stroke structures, with moderate blur and isolated artifacts. For Fashion-MNIST, garment silhouettes, brightness, and broad texture are largely recovered, while finer details, such as clothing prints, are smoothed. In addition, mono-SIM and joint constructive configurations give visually comparable reconstructions with both semantic targets, and disjoint symmetric alignment is similarly effective for the Linear target. With PPFE, the disjoint symmetric configuration yields visibly less faithful outputs: some MNIST digits (notably $0$, $4$, $8$, and $9$) show noisier strokes, and Fashion-MNIST images appear less sharp. A plausible contributing factor is the smaller number of PPFE prototypes (at most $128$, versus $2000$ for the other designs), which limits the rank and descriptive richness of the resulting target operator. Overall, these results extend the evidence for OTA semantic alignment beyond classification to image reconstruction with independently trained models.

% ---------------- Supplementary references (kept separate from the main list) ----------------
\begingroup
\renewcommand{\refname}{Supplementary References}

\endgroup

\end{document}